\documentclass[10pt]{article}
\pdfoutput=1

\usepackage{arxiv}  

\usepackage{times}
\usepackage[hyphens]{url}
\usepackage{hyperref}
\PassOptionsToPackage{hyphens}{url}\usepackage{hyperref}
\hypersetup{pdfpagemode=UseNone} 
\hypersetup{colorlinks, urlcolor={blue}}
\usepackage{multicol}
\usepackage{caption}

\usepackage[section]{placeins}
\usepackage[fleqn]{amsmath}

\usepackage{tikz}
\usetikzlibrary{positioning}
\usetikzlibrary{shapes}
\usepackage{iftex}
\ifPDFTeX
\else
   \usepackage{svg} 
   
\fi

\usepackage{natbib}

\newcommand{\myvec}[1]{\boldsymbol{#1}}

\usepackage{orcidlink}
\newcommand{\orcid}[1]{\orcidlink{#1}}

\title{Using Small Domain Estimation to obtain better retrospective Age Period Cohort insights}

\author{
Koen Simons \orcid{0000-0002-6534-2277}\\
Epidemiology and Biostatistics \\
Melbourne School of Population and Global Health \\
University of Melbourne, AU \\
\texttt{koen.simons@unimelb.edu.au}
\And Rebecca Bentley \orcid{0000-0003-3334-7353}\\
Centre for Health Equity \\
Melbourne School of Population and Global Health \\
University of Melbourne, AU \\
\texttt{brj@unimelb.edu.au}
\And Lyle Gurrin \orcid{0000-0001-7052-1969}\\
Epidemiology and Biostatistics \\
Melbourne School of Population and Global Health \\
University of Melbourne, AU \\
\texttt{lgurrin@unimelb.edu.au}
}

\begin{document}
\maketitle

\begin{abstract}
Recent changes in housing costs relative to income are likely to affect people's propensity to Housing Affordability Stress (HAS), which is known to have a detrimental  effect on a range of health outcomes. The magnitude of these effects may vary between subgroups of the population, in particular across age groups. Estimating these effect sizes from longitudinal data requires Small Domain Estimation (SDE) as available data is generally limited to small sample sizes. In this paper we develop the rationale for smoothing-based SDE using two case studies: (1) transitions into and out of HAS and (2) the mental health effect associated with HAS. We apply cross-validation to assess the relative performance of multiple SDE methods and discuss how SDE can be embedded into g-computation for causal inference.
\end{abstract}


\section{Introduction}

The number of people experiencing housing affordability stress (HAS) is increasing in many high-income countries due to the rapid rise in the past 10 years of the cost of housing relative to incomes \citep{taylor2017secular}. At the same time, a growing number of studies in a variety of settings report associations between  HAS and mental health \citep{bentley2011association, mason2013housing, bentley2016housing, mari2017housing}. Many of these studies have considered changes in people's housing situation over time to assess concordant changes in their health status. On the whole, there is evidence that HAS has a short term negative effect on mental health. The increasing prevalence of HAS combined with the known relationship between housing affordability and the risk of poor mental health is, potentially, an important determinant of population health. It may provide an indicator at the population level of the ``health" of housing within a given society or country, in this case Australia. Here we attempt a big-picture view of this relationship by looking across time and tracking age-based cohorts, seeking to describe the impact of the cost of housing over time on population well-being.

The experience of being unable to afford the cost of housing is an obvious source of stress. Despite this, housing affordability stress is difficult to define and multiple operational definitions exist \citep{nepal2010measuring}. A definition that has been widely used in research and by national data collection agencies is the so-called 30/40 rule, where a household is considered to be in housing affordability stress if it spends at least 30\% of its gross income on housing costs and is in the bottom 40\% of the national equivalized disposable income distribution \citep{trewin2002measuring}. For a household to ``exit" from unaffordable housing requires an opportunity to increase income and/or an opportunity to decrease housing-related expenses, such as through improved employment status, moving house or receiving government subsidies. Clearly, the abundance or scarcity of such opportunities depends on both personal and market characteristics. For example, education and experience are personal factors that influence employment opportunities, whereas house prices differ strongly by location. Financial transactions within families (such as gifts or bequests) may alter these probabilities on the individual level, and expectations about wealth and attitudes to the management of finances will be influenced by parents	 and the wider family environment \citep{cigdem2017intergenerational}. While transfers are generally within-family, expectations reflect comparisons with peers and/or previous generations; they are not limited to the individual's circumstances. The probability of exiting from HAS is, therefore, likely to be dependent on region, age and social stratum. Since demographics, policy and market conditions naturally change over time, there is possibly temporal variation of the exit probability. Most of these factors would also be expected to influence the probability of exposure to or ``entry" into HAS. At the population level, these combined influences result in age-, time-, region- and social stratum-specific differences in the incidence of and emergence from HAS. Identifying and understanding these differences may lead to more efficient policies by targeting interventions to lift people out of HAS. Similarly, identifying differences in the rates of entry into HAS enables targetted preventive measures. 

There are complexities in estimating effects of HAS on health outcomes.  Ideally stratum-specific entry and exit rates would be estimated directly from longitudinal data where HAS status is recorded at two or more times. Census information may also provide a large or exhaustive sample that could be used to estimate the prevalence of HAS. Such data are, however, unlikely to be set up for longitudinal analysis, so entry and exit probabilities cannot be obtained without further assumptions. Similarly, estimation of stratum-specific effects of HAS requires that information on health outcomes is linked to exposure data. While the frequency of the creation of record-linked data resources continue to improve, concerns for privacy may limit access to the resulting data.

One feature that these data platforms share is that sufficiently rich data for the analyses we propose are likely to be limited to small sample sizes. Moreover, small sample sizes lead to even smaller sample sizes per stratum. These very small sample sizes provide little information on stratum-specific rates and effects. Small Domain Estimation (SDE) or Small Area Estimation (SAE) aims to overcome this problem and can be used to obtain efficient estimates of HAS incidence and prevalence. Here, small refers to the sample size of an individual stratum, as opposed to the geographical area or the total population size of a stratum. Domain refers to the level of aggregation that is the target of estimation; a domain may be composed of multiple strata. Efficiency for domain-specific estimates is gained by including information about the structure of the data or through a trade-off between the bias and the variance of estimation. \cite{rao2015small} and \cite{pfeffermann2013new} describe various statistical methods to obtain small domain estimates. \cite{rahman2013simulating} evaluated the use of SAE to obtain efficient estimates of HAS prevalence from survey data. Their findings indicate that there is large geographical variation in HAS: ``Almost two-thirds of these households are located in statistical local areas (SLAs) in eight capital cities, and a large number of them are in Sydney and Melbourne."

In this paper, we focus primarily on variation in entry and exit rates by age, birth-year and calendar-year and do not consider explicitly spatial variation or produce sex-specific estimates. First, we introduce three traditional estimators to discuss the bias-variance trade-off: direct estimation, complete pooling and partial pooling. Second, we define an estimator based on tensor smoothing splines that exploits the relative similarity between adjacent domains, and we discuss the strengths and weaknesses by applying Cross-Validation on an example dataset. Third, we consider the effect of HAS on changes in Mental Health (MH) and the modification of these changes by age, birth-year and calendar year. We illustrate how the different Small Domain Estimators can be used to estimate both the main effect of HAS on Mental Health as well as for estimating and visualising the heterogeneity of effect. We discuss the limitations of these methods in the context of Age-Period-Cohort data where, for identification, an arbitrary constraint must be imposed. We argue, however, that this does not limit interpretability of the findings. We conclude that tensor splines and other SDE approaches are useful for detecting trends and/or hot spots as well as for generating hypotheses. If backed by appropriately specified Directed Acyclic Graphs (DAG), the ability to explicitly allow for heterogeneity of effect may be advantageous when employing a potential outcomes approach to causal inference.

\section{Data}

Since 2001, the Melbourne Institute of Applied Economic and Social Research is annually collecting data for the longitudinal Household, Income and Labour Dynamics in Australia Survey (HILDA). At baseline, 7682 households were interviewed and 19914 individuals were included. As is common in such studies, participants drop out for various reasons. In 2011, a `top-up' sample was added, consisting of 2153 new households or 5451 new individuals.

HILDA data has previously been used to investigate HAS \citep{borrowman2017long} and the effects of housing conditions on mental health \citep{bentley2018impact,singh2019financial}. For this study, we included, for each calendar-year, all individuals who were between 25 and 64 years old and who had moved out of their parents house. Individuals with missing data on household equivalised income or housing-related costs were discarded for the affected time periods. Similarly, analysis of changes in mental health, SF36, was restricted to available data.

\section{Entry and Exit Rates}

\subsection{SDE, partial pooling and smoothing splines}

Entry into and exit from Housing Affordability Stress can be seen as a two compartment model: the population not in HAS, who are at risk of entering into HAS, and the population in HAS, who may exit. We assume that the two transition processes are first order Markov processes, specifically:
\begin{align*}
p_{\mathrm{entry}} & = p \left(Y_{it} = 1 \vert Y_{it-1} = 0,  Y_{it-2}, Y_{it-3}, \ldots{}, A_{it}, A_{it-1}, A_{it-2}, \ldots{} \right) \\
& = p \left(Y_{it} = 1 \vert Y_{it-1} = 0, A_{it} \right) \\
p_{\mathrm{exit}} & = p \left(Y_{it} = 0 \vert Y_{it-1} = 1, A_{it} \right) 
\end{align*}
Wherein $Y_{it} = 1$ indicates that individual $i$ experienced HAS in year $t$ and $A_{it}$ represent covariates such as age. 

Given a sample $S$ of observations $(t, a_t, y_{t}, y_{t-1})$, unbiased estimates of the conditional entry and exit probabilities can be obtained directly from the sample averages:
\begin{align*}
\hat{p}_{\mathrm{exit}}^{d}(a,t) & = \sum_{i : Y_{it-1} = 1, A_{it} = a} Y_{it} / n_{at} 
& n_{at} = \# \lbrace i \in S : Y_{iT-1} = 1, A_{iT}=a, T = t \rbrace
\end{align*}

\begin{figure}
\centering
\includegraphics[width=0.45\textwidth]{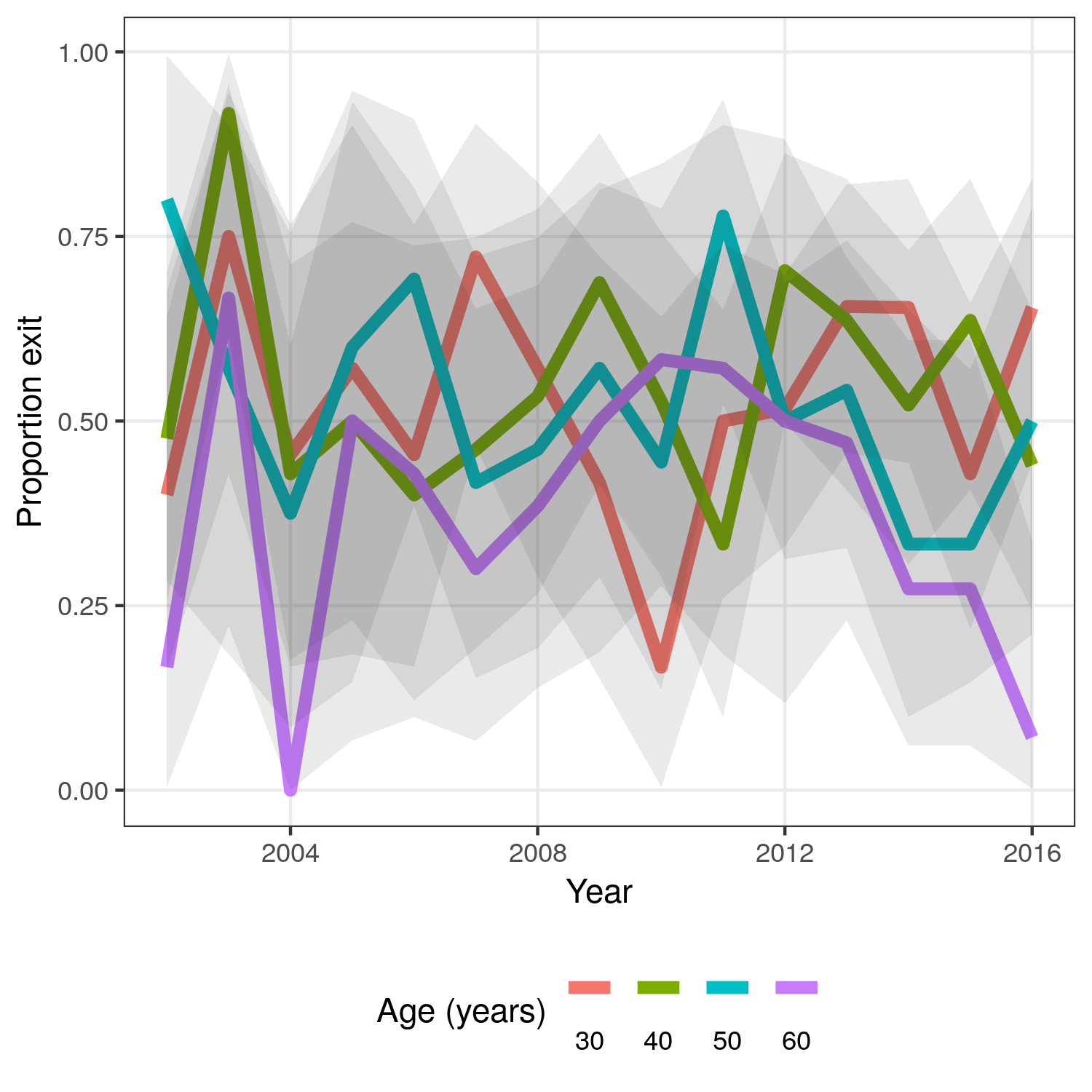} 
\caption{Probability of exit from unaffordable housing estimated using direct estimation from the sample, without pooling. For ease of visualisation only ages 30, 40, 50 and 60 are shown.\label{fig.exitnopool}}
\end{figure}

However, the variance of the \emph{direct} estimators $\hat{p}^{d}(a,t)$ is $O(1/n_{at})$ and will be large for small $n$ as is illustrated in Figure \ref{fig.exitnopool} for the exit probability, which suffers smaller denominators as the number of individuals with HAS is a small fraction of the population as shown in Figure \ref{fig.atrisk} in appendix.

A straightforward way to reduce the variance is to combine the observations from multiple domains and calculate a \emph{pooled} estimate. The \emph{completely pooled} estimator is defined as:
\begin{align*}
\hat{p}_{\mathrm{exit}}^{c}(a,t) & = \sum_{i,T : Y_{iT-1} = 1} Y_{iT} / n 
& n = \sum_{A,T} n_{AT} 
\end{align*}

This estimate has lower variance, however, the estimator is biased unless the transition probability is homogeneous over all domains. Moreover, the magnitude of the bias depends on the heterogeneity and is unknown. It is possible to define alternative estimators with a different trade-off between bias and variance, e.g. by combining (aggregating) domains into a set of super-domains. Considering the previously introduced estimators, a compromise between the unbiased direct estimator and the biased completely pooled estimator can be obtained by weighting the estimators:
\begin{align*}
\hat{p}_{\mathrm{exit}}^{w}(a,t) & = w_{at} \hat{p}_{\mathrm{exit}}^{c}(a,t) + (1-w_{at}) \hat{p}_{\mathrm{exit}}^{d}(a,t)
\end{align*}

For each domain $(a,t)$, the weighted estimate is between the direct and completely pooled estimates. Hence, they can be seen as shrinkage estimators that pull the direct estimators towards an overall mean and reduce the heterogeneity of the estimates without requiring it to be zero. This produces estimates that have a variance between the completely pooled and the direct estimates and bias must also be between the two. Naturally, the amount of bias and variance depend on the choice of the domain-specific weights. There are multiple ways to obtain suitable weights $w_{at}$. One option is to rewrite the problem as a (Bayesian) hierarchical model such that there is an overall mean odds ($p_o$) but domain-specific odds vary around this mean. Working on the log-odds scale, a typical model is:
\begin{align*}
Y_{it} \vert a & \sim \mathrm{Binomial}(p_{a,t}) \\
\mathrm{logit} (p_{a,t}) & = \pi_{at}  & u_{at} \sim N(0,\sigma) \\
\pi_{at} & = \pi_o + u_{at}
\end{align*}

In this approach, the amount of shrinkage can be controlled indirectly by choosing a prior for $\sigma$ that puts more or less weight on values close to 0. Together with the data, this results in a posterior distribution for $\sigma$ and for both the overall mean probability $p_o$ and the domain-specific transition probabilities $p_{at}$. The estimates for domains with lower sample sizes, relative to the average sample size, will be shrunk more than those with more observations. In addition, the weighted estimates wherefore the respective direct estimates are relatively further away from the overall mean, will be shrunk more than those that are closer to the overall mean. Besides shrinking, this property is also referred to as `regularization', as it shrinks outlying values towards the mean; `borrowing strength' or \emph{`partial pooling'} as it augments the domain-specific sample with observations from other domains in such a way that the amount of pooling or borrowing depends on the relative sample size.

\begin{figure}
\centering
\includegraphics[width=0.45\textwidth]{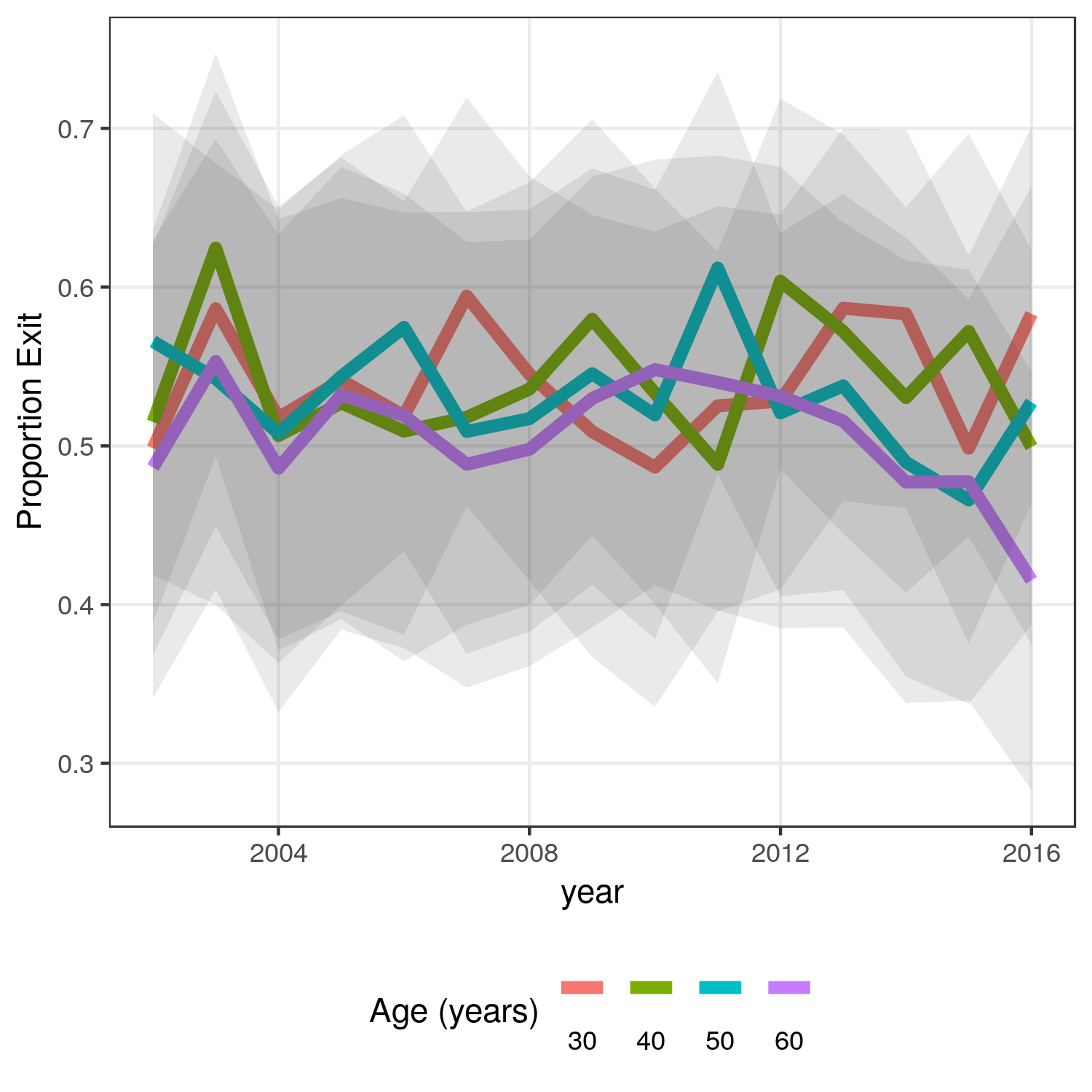}  \quad \includegraphics[width=0.45\textwidth]{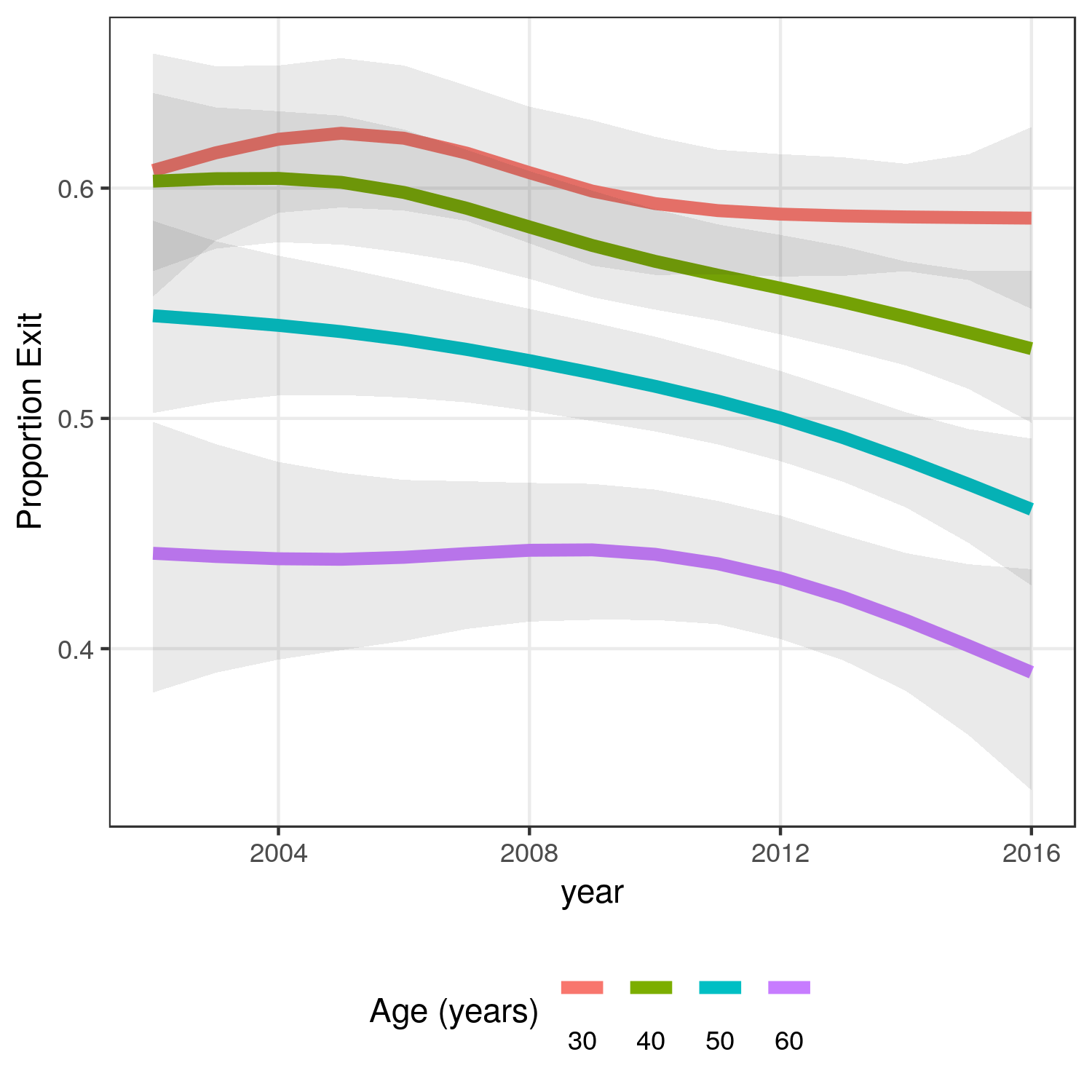} 
\caption{Probability of exit from unaffordable housing estimated using small domain estimation. Left: estimates obtained from a partial pooling model with iid error. Right: estimates obtained using a tensor spline basis. For ease of visualisation only ages 30, 40, 50 and 60 are shown. \label{fig.exitsde}}
\end{figure}

Throughout the remainder or this manuscript we will refer to such estimates as partially pooled estimates. Results of partial pooling for the exit probabilities are visualized in the left panel of Figure \ref{fig.exitsde}. As expected, partial pooling results in estimates that are less heterogeneous than the direct estimates presented earlier. Comparison with Figure \ref{fig.exitnopool} reveals gains in \emph{stated} precision, however, credible intervals remain wide and no patterns across age or time are revealed.

Nonetheless, such patterns are intuitively expected: aggregate changes in market conditions and in demographics are often slow. Furthermore, conditions for a 30-year-old person will be more similar to conditions for a 31-year-old person than those for a 60 year old person. It is sensible to expect that the entry and exit probabilities vary smoothly over time and age:
\begin{align*}
Y_{it} \vert a & \sim \mathrm{Binomial}(p_{a,t}) \\
\mathrm{logit} (p_{a,t}) & = s(a,t)
\end{align*}
Wherein $s$ is a \emph{smooth} 2d function. 

Options for estimating $s$ include 2d kernels, thin plate smoothing splines and tensor splines. To distinguish the smooth estimators from partially pooled estimators, consider the naive kernel that assigns weight $w_{at} = 1$ if $\vert A - a\vert < 5$ \& $\vert t - T \vert < 5$ and $w_{at} = 0$ elsewhere. Define the naive smooth estimator as:
\begin{align*}
\hat{p}_{\mathrm{exit}}^{n}(a,t) & = \sum_{i,T : Y_{iT-1} = 1}^{\vert A_{iT} - a\vert < 5, \vert T - t \vert < 5 } Y_{iT} / n_{at} \\
n_{at} & = \# \lbrace (T, A_{iT}, Y_{iT}, Y_{iT-1}) \in S : Y_{iT-1} = 1, \vert A_{iT} - a\vert < 5, \vert T - t \vert < 5 \rbrace
\end{align*}

This naive smooth estimator uses overlapping age-period windows of ten years by ten years and completely pools all the data within each window while ignores all data outside of the window. Using a moving window yields an average absolute difference between estimates for two adjacent domains that will be lower than when direct estimation is used. In other words, the estimates are less \emph{wiggly}\footnote{For twice differentiable functions $f(x)$, wiggle can be defined as $\int \vert \partial^{2} f(x)/\partial x^{2} \vert dx$, ie a line with constant slope has zero wiggle}. This can be further improved upon by using a different kernel such that more weight is given to observations closer to the centre of the window, optimising the width of the window, allowing the window's width and height to be different, etc. Indeed, the naive kernel has multiple deficiencies, including poor performance at the edges of the data: the estimate for the youngest age $a_{\mathrm{min}}$ will include information from observed ages $a_{\mathrm{min}}$ to $a_{\mathrm{min}} + 5$ and if there is a monotonous trend in the exit probability over age, then the estimate will be closer than the estimate for age $a_{\mathrm{min}} + 2.5$; the estimate is biased. Complete and partially pooled estimates will also be biased when a trend exists as they pool all data towards the global mean regardless of the existence of trends.

Instead, a linear model could be adopted to account for the possibility of linear trends, quadratic and higher order terms can be included to allow for more complex patterns. It is well known that using piecewise polynomials such as cubic b-splines can be used to perfectly fit data as long as enough knots are included in the basis functions; a saturated tensor b-spline model provides the same estimates a the direct estimator. By applying a penalty to the amount of wiggle, p-splines \citep{eilers1996flexible} shrink the coefficients that describe the response surface towards a flatter but not necessarily more horizontal surface. Hence, tensor p-splines are a useful method for estimating 2d smooth functions such as $s(a,t)$. Other options include thin plate smoothing splines, radial base splines and methods that implicitly estimate the correlation between parameters for adjacent cells. \cite{opsomer2008non} specifically used radial bases of p-splines to create small domain estimates; the Besag-York-Mollie (BYM) model includes random effects that are spatially correlated, relying either on an adjacency matrix or on a distance metric \citep{besag1991bayesian}. Assuming there is zero spatial correlation reduces the BYM model to partial pooling.

Despite similarities in aims, these options are not equivalent and the relative merits are often studied by simulations, e.g. \cite{chambers2009borrowing}. For this application, we recommend tensor smoothing splines as there is no reason to expect the response surface to be isotropic: there is no prior reason to enforce that age 30 is equally similar to age 31 as 2002 is similar to 2003. Tensor splines are invariant to the choice of scales \citep{wood2013straightforward}, whereas thin plates are isotropic on the chosen scale and conditional autoregressive models are similarly dependent on the relative scaling of age and calendar time.

\subsection{Evaluating performance}

\begin{figure}
\centering
\includegraphics[width=0.45\textwidth]{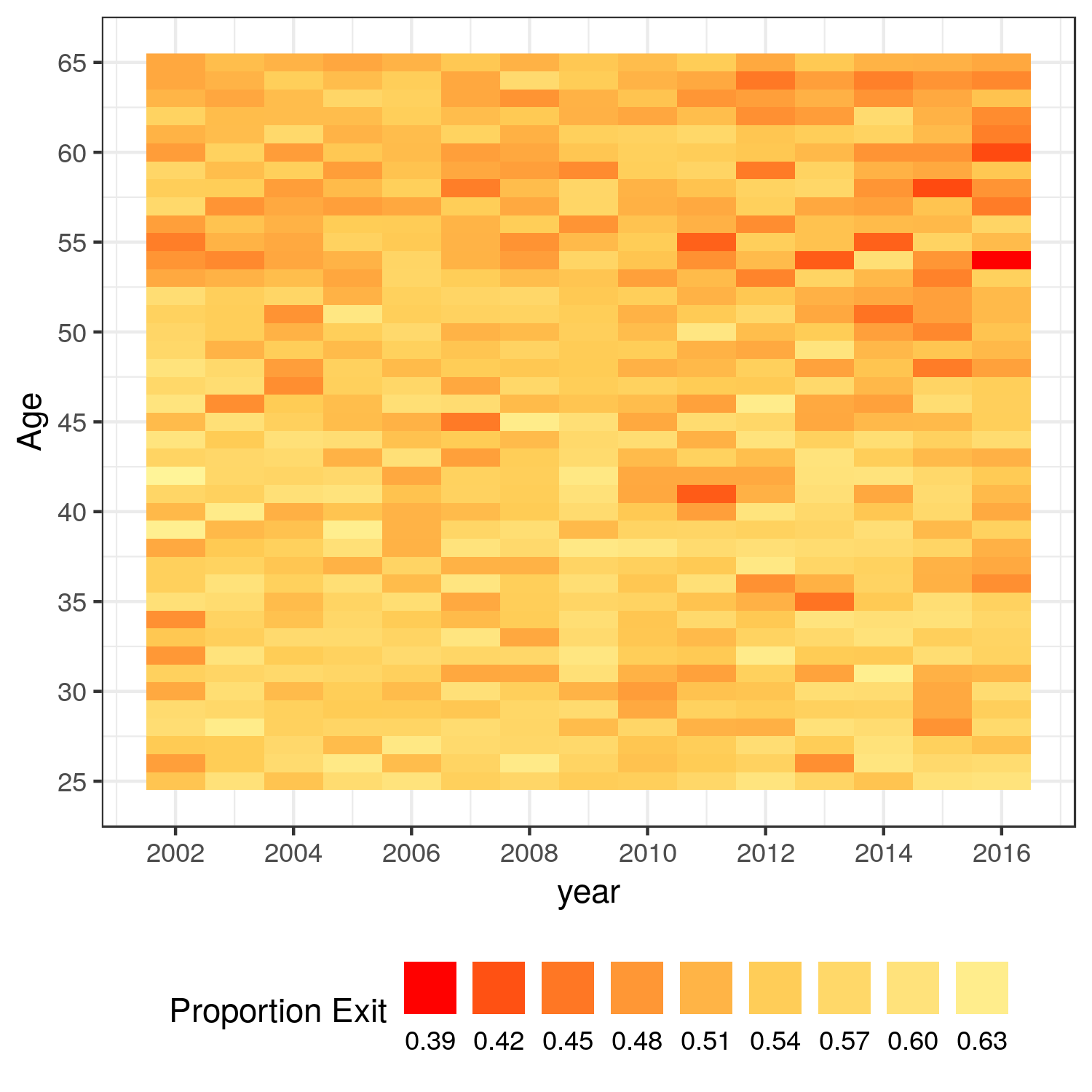} \quad \includegraphics[width=0.45\textwidth]{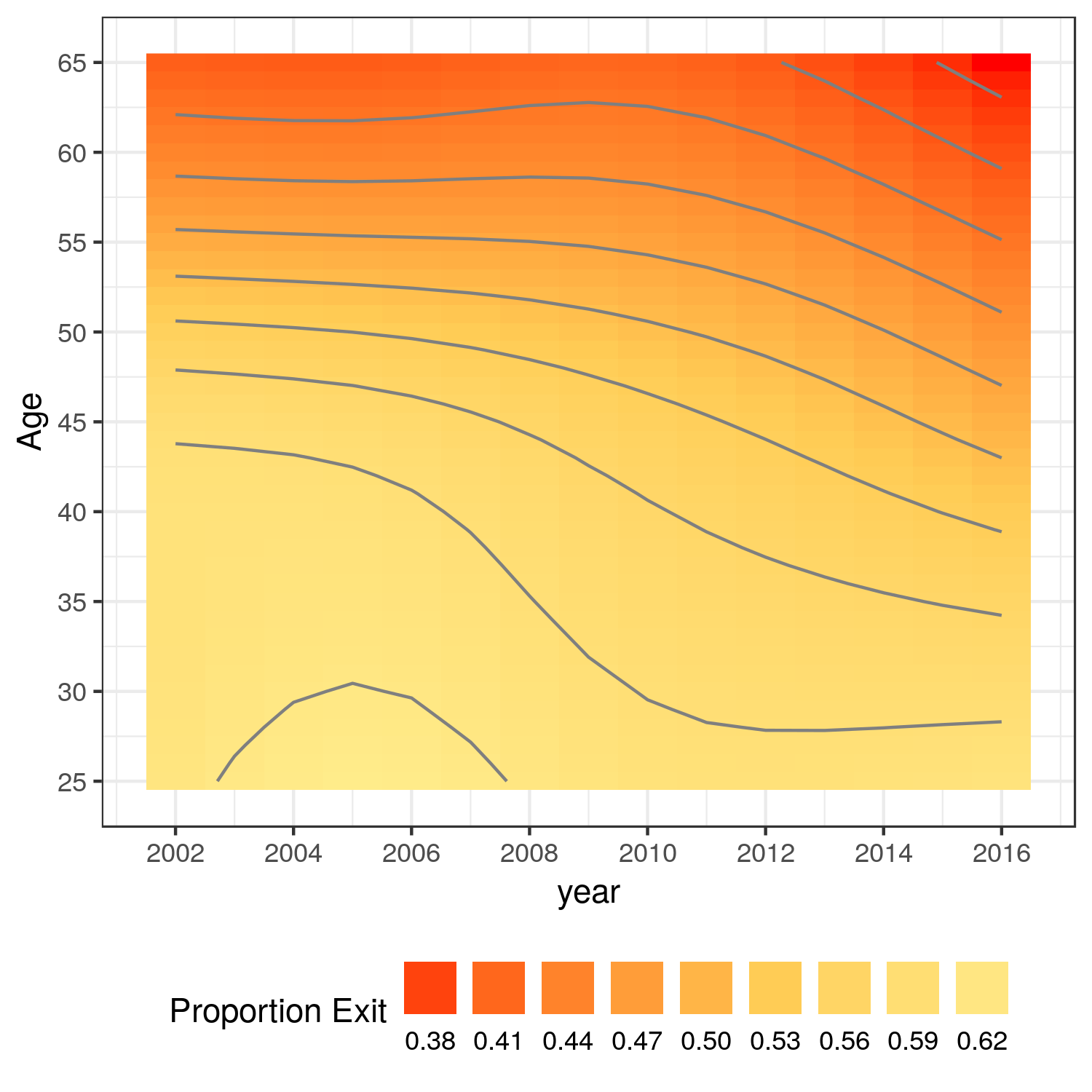}
\caption{Heatmaps of posterior means for the probability of exit by APC. Left: map obtained from a partial pooling model with iid error. Right: map obtained using a tensor spline basis.\label{fig.heatmapexit}}
\end{figure}

We used the R \citep{r2018R} package {\tt brms} \citep{burkner2017brms} to estimate a smoothing tensor spline response surface for the exit probabilities. The right panel of Figure \ref{fig.exitsde} shows that the (stated) precision is greatly increased and two trends are revealed: the probability of exit from HAS decreases with age and has, for the most part, been declining over the study period. This figure is restricted to four one-year age-groups for visual clarity but has the advantage that the credible intervals are included. The companion Figure \ref{fig.heatmapexit} shows the patterns across all age-calendaryear combinations but without indication of uncertainty. 

In Figures \ref{fig.exitsde} and \ref{fig.heatmapexit}, based on the HILDA data, show that only tensor splines yield visually clear gradients. Such gradients facilitate the generation of hypotheses that explain the apparent differences across domains. Moreover, Figure \ref{fig.exitsde} demonstrates that tensor smoothing splines can result in higher \emph{stated} precision compared to direct and partially pooled estimates. However, complete pooling provides even more optimistic \emph{stated} precision: the pooled estimate is 46\% with 95\% confidence interval ranging 45\% to 47\%. Figure \ref{fig.exitsde} shows that the other methods produce point-wise intervals that are at least four times wider. Considering only stated precision, tensor splines remain second to complete pooling. However, it is known that complete pooling is biased if effects are indeed heterogeneous. Instead of relying on the width of confidence or credible intervals, model fit can be assessed more formally by calculating the \emph{expected log posterior predictive density} of the observed data, given model $m$:
\begin{align*}
\mathrm{elpd}^{(m)} & = \sum_i \mathrm{log}p(y_i \vert \myvec{y}, m) \\
& =  \sum_i \int \mathrm{log} \left( p(y_i \vert \hat{\pi}_{i}^{(m)}) \right) p ( \hat{\myvec{\pi}}^{(m)} \vert \myvec{y}, m )  d \hat{\myvec{\pi}}^{(m)} \\
& \sim  \sum_i \int \mathrm{log} \left( p(y_i \vert \hat{\pi}_{i}^{(m)}) \right) p ( \myvec{y} \vert \hat{\myvec{\pi}}^{(m)} , m )  p(\myvec{\hat{\pi}}^{(m)}) d \hat{\myvec{\pi}}^{(m)}
\end{align*}
Through the rest of the paper we will omit the dependency on the model $m$. The estimated variability over the domains and the average stated precision are calculated using: 
\begin{align*}
V(E\left[\hat{\pi}_{at}\right]) & \sim \sum_{at} \int \left( \hat{\pi}_{at} - \hat{\pi}_{o} \right)^2 p(\myvec{y} \vert \myvec{\hat{\pi}}) p(\myvec{\hat{\pi}}) d\myvec{\hat{\pi}}\\
E\left[V(\hat{\pi}_{at})\right] & \sim \sum_{at} \int \left( \hat{\pi}_{at} - \bar{\pi}_{at} \right)^2 p(\myvec{y} \vert \myvec{\hat{\pi}}) p(\myvec{\hat{\pi}}) d\myvec{\hat{\pi}}
\end{align*}
Wherein $\pi_o$ represents the mean log odds over all observations and $\bar{\pi}_{at}$ is the domain-specific posterior mean:
\begin{align*}
\hat{\pi}_{o} & = \frac{1}{n} \sum_i \hat{\pi}_{i} \\
\bar{\pi}_{at} & \sim \int \hat{\pi}_{at} p(\myvec{y} \vert \myvec{\hat{\pi}}) p(\myvec{\hat{\pi}}) d\myvec{\hat{\pi}}
\end{align*}

As defined above, elpd provides a measure of fit and does not penalise the complexity of the model. Therefore, it is likely to favour models that over-fit the data. In order to more objectively evaluate the performance of the estimators, we used 5-fold Cross-Validation (CV) stratified by age-calendar-year; we repeatedly held out approximately\footnote{Due to the small sample sizes it is not always possible to withhold exactly 20\% of the data in each domain, e.g. if there are 6 observations in a age-period combination, one of the folds must include two observations and the other four folds contain one observation each.} 20\% of the sample data from each domain and generated posterior predictions for these test datasets based on the remaining 80\% of the data as training data. The above definition of expected log posterior density is trivially extended to new data. Using the same folds for each model, it is possible to calculate the pairwise differences in expected log posterior predictive density:
\begin{align*}
\mathrm{elpd}_i^{(m)} & \sim  \int \mathrm{log}p(y_i \vert \myvec{\hat{\pi}}) p(\myvec{y}^{\mathrm{train}} \vert \myvec{\hat{\pi}}) p(\myvec{\hat{\pi}}) d\myvec{\hat{\pi}} \\
\Delta^{(m,w)} \mathrm{elpd} & = \sum_i ( \mathrm{elpd}_i^{(m)} - \mathrm{elpd}_i^{(w)}  )
\end{align*}

\begin{table}[bt]
\caption{\label{tab.elpd.exit} Comparison of out of sample predictive performance of estimates of age-calendar-year-specific probabilities of exit from unaffordable housing, using 5-fold Cross-validation stratified by age-calendar-year.}
\begin{center}
\begin{tabular}{lrrrrrlrl}
\hline
Estimator & \multicolumn{4}{c}{Whole sample} & \multicolumn{4}{c}{5-fold CV} \\ \cline{2-5} \cline{6-9}
 & elpd & $\Delta$ elpd & $V(E\left[\hat{\pi}_{at}\right])$* & $E\left[V(\hat{\pi}_{at})\right]$* & elpd & sd & $\Delta$ elpd & sd  \\
\hline
 Complete pooling: $\pi_{at} = \pi_c$ & -6209.7 & -112.2 & 0.0 & 0.4 & -6210.0 & 7.2 & -29.7 & 7.4 \\ 
  Partial pooling: $\pi_{at} = \pi_o + u_{at}$ & -6097.4 &  & 31.2 & 74.6 & -6180.4 & 10.3 &  &  \\ 
  Tensor spline: $\pi_{at} = \pi_o + s(at)$ & -6125.0 & -27.5 & 74.1 & 4.5 & -6129.8 & 14.5 & 50.5 & 11.8 \\ 
  
\hline
 Complete pooling**: $\pi_{at} = \pi_c$ & -6209.7 & -112.3 & 0.0 & 0.4 & -6210.7 & 7.2 & -30.3 & 7.4 \\ 
  Direct estimator**: $\pi_{at} = \pi_{at}$ & -6052.6 & 44.9 & 369.9 & 279.8 & -6775.4 & 32.7 & -595.0 & 25.2 \\ 
  
\hline
\end{tabular}
\end{center}
\footnotesize{* Rescaled by 1000 \\
** Partial pooling and tensor spline estimators were obtained using the package {\tt brms}, however, the direct estimator was not. To account for possible numeric deficiencies, the completely pooled estimator was fit both with and without brms.}
\end{table}

\subsection{Results}

Table \ref{tab.elpd.exit} lists the expected log posterior predictive density calculated both with the whole sample and with the cross-validation procedure. Looking at within-sample assessment, the direct estimator has an expected log posterior density of -6052.6, which is higher than the elpd of the other methods. The within-sample assessment of performance is often too optimistic. In particular, the direct estimator performs substantially worse in a ``new'' sample: the elpd drops to -6775.4. In contrast, the completely pooled estimator has near identical performance within and out of sample, elpd -6209.7 respectively -6210. A small decrease is to be expected in cross-validation as the procedure uses less data to fit the model. As expected, partial pooling is a favourable compromise between complete pooling and direct estimation; with elpd -6180.4 under CV it performs better than the two estimators it is based on. However, the largest elpd under CV for this dataset, -6129.8, is obtained by the tensor spline-based estimates. Its within-sample elpd is close to that obtained under CV, whereas partial pooling is more optimistic within-sample.

Partial pooling and tensor splines outperform direct estimation and complete pooling on out of sample assessment. This improvement can be attributed to a favourable bias-variance trade. This trade is possible due to variation across domains in true transition probabilities; if there is no such variation then the completely pooled estimator is unbiased and has minimal variance. However, Table \ref{tab.elpd.exit} shows that the direct, partially pooled and tensor spline estimates vary across domains. For this dataset, tensor splines net twice as much variation in estimated probabilities as partial pooling (0.0741 respectively 0.0312) but not nearly as much variation as the direct estimator (0.3699). Simultaneously, tensor splines have the second highest precision: the variability of the posterior, averaged over all domains, is 0.0045 for tensor splines, 0.0004 for complete pooling and much higher for partial pooling and direct estimation. If there truly is heterogeneity, the reduced heterogeneity in the estimates implies bias. This bias is traded for reduced variability of the posterior compared to the direct estimator. In this sample, the trade-off provides a net benefit for both partial pooling and tensor splines but the exchange is much more favourable for tensor splines: less bias is taken for a larger reduction in variance.\footnote{For partial pooling the trade-off can be further optimised, by manipulating the weights $w_{at}$ directly or through the prior on $\sigma$. For tensor splines the trade-off can also be optimised by changing the penalty parameter (or by altering the number of knots used to construct the b-spline bases). For both methods, a set of solutions can be defined and indexed by one or more `tuning parameters'. Choosing the optimal tuning parameter is a direct trade of bias for variance within a method- and data specific set of solutions. Clearly, tensor splines have a favourable exchange rate here.}

To understand how tensor splines attain higher efficiency, we repeat the Cross-Validation with a different set-up. Instead of sampling within each domain, it is possible to perform cross-validation by repeatedly holding out entire domains, e.g. excluding ages (corresponding to rows in Figure \ref{fig.heatmapexit}), calendar-years (columns) or birth-years (diagonals). Table \ref{tab.elpd.exit.wave} shows the results for leaving out calendar-years: the performance of complete pooling remains the same as under the stratified CV scenario, however, partial pooling now performs on par with complete pooling, while it was superior previously. Tensor spline smoothing remains superior in the new scenario but the difference in elpd within-sample and under cross-validation has changed from -4.8 (= -6129.8 + 6125) to -13.7 (= -6138.7 + 6125). These results illustrate that tensor spline smoothing combines information from the target domain with neighbouring domains to efficiently estimate domain-specific exit probabilities. If only neighbour domains are available, the procedure still works but the accuracy is affected. In contrast, partial pooling combines information from the target domain and the overall mean. If no observations from the target domain are available, the predictions effectively use information from the overall mean only and hence the predicted values are very similar to those derived by complete pooling. If there are trends in over the domains, shrinking the estimate towards the overall mean may generate more bias than shrinking the estimate towards the nearest neighbours' mean.

\begin{table}[bt]
\caption{\label{tab.elpd.exit.wave} Comparison of out of sample predictive performance of estimates of age-calendaryear-specific probabilities of exit from unaffordable housing, using 5-fold Cross-validation wherein entire calendaryears were left out.}
\begin{center}
\begin{tabular}{lrrrrrlrl}
\hline
Estimator & \multicolumn{4}{c}{Whole sample} & \multicolumn{4}{c}{5-fold CV} \\ \cline{2-5} \cline{6-9}
 & elpd & $\Delta$ elpd & $V(E\left[\hat{\pi}_{at}\right])$* & $E\left[V(\hat{\pi}_{at})\right]$* & elpd & sd & $\Delta$ elpd & sd  \\
\hline
 Complete pooling: $\pi_{at} = \pi_c$ & -6209.7 & -112.2 & 0.0 & 0.4 & -6211.3 & 7.3 & -0.3 & 0.5 \\ 
  Partial pooling: $\pi_{at} = \pi_o + u_{at}$ & -6097.4 &  & 31.2 & 74.6 & -6211.0 & 6.8 &  &  \\ 
  Tensor spline: $\pi_{at} = \pi_o + s(at)$ & -6125.0 & -27.5 & 74.1 & 4.5 & -6138.7 & 15.1 & 72.3 & 13.1 \\ 
  
\hline
\end{tabular}
\end{center}
\footnotesize{* Rescaled by 1000}
\end{table}

\section{Effect of HAS on Mental Health}

\subsection{Modelling choices}

The combination of demographic change with changing market conditions may lead to changes in expectations and modify the mental health effects associated with HAS. To investigate this, we again assume that mental health scores are the result of a first order Markov Process, specifically:
\begin{align*}
p\left(Y_{it} = y_{it} \vert Y_{it-1}, Y_{it-2}, \ldots{}, A_{it}, A_{it-1} \ldots{} \right) & = p\left(Y_{it} = y_{it} \vert Y_{it-1},  A_{it} \right)
\end{align*}

The distribution of SF36 is skewed and it is reasonable to expect that, without external influences, there is no change in Mental health: $E[Y_{it}\vert Y_{it-1}] = Y_{it-1}$. Hence, we opt to model the first differences:
\begin{align*}
\Delta Y_{it} & = Y_{it} - Y_{it-1}
\end{align*}

The distribution of first differences is symmetrical, however, there is evidence of conditional heterogeneity and regression to the mean (Figure \ref{fig.sf36} in appendix). To accommodate this we fitted the following tensor spline model:
\begin{align*}
\left\{ \begin{array}{ll}
\Delta Y_{it} & \sim N \left(\mu_{it}, \sigma_{it} \right) \\
\mu_{it} & = \alpha + \beta Y_{it-1} + s_1(A_{it},t) \\
\mathrm{log}\sigma_{it} & = s_h(Y_{it-1})
\end{array} \right. &
\end{align*}
Wherein $s_1$ is a smooth 2d function modelled using tensor splines and $s_h$ is a 1d smooth function. For convenience, we use fully-parametric cubic b-splines with 5 degrees of freedom for $s_h$. Similar to the models for entry and exit, a comparison partial pooling model is defined by replacing $s_1$ by $u_{at} \sim N(0,\sigma_d)$. 

\begin{figure}
\centering
\begin{tikzpicture}[
	align=center,node distance=2cm,
	squarednode/.style={rectangle, draw=black, rounded corners, thick, minimum size=5mm},
	scale=0.45
	]
	\input{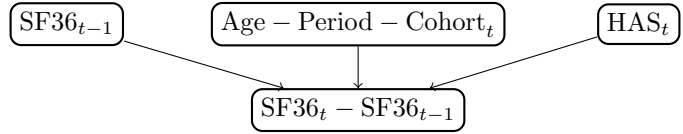}
\end{tikzpicture}
\caption{Direct Acyclic Graph for the evolution of Mental Health influenced by Housing Affordability Stress and an age-period-cohort effect.\label{fig.DAG.simple}}
\end{figure}

These models allow for an age and calendar-year effect on the change in Mental Health but do not include HAS. We contrast these models with two alternatives. The first includes an effect of HAS ($M$) on changes in Mental Health ($\Delta Y$), the second allows for the effect of HAS to be modified by age and calendar-year:
\begin{align*}
\left\{ \begin{array}{ll}
\Delta Y_{it} & \sim N \left(\mu_{it}, \sigma_{it} \right) \\
\mu_{it} & = \alpha + \beta_1 Y_{it-1} + s_1(A_{it-1},t) + \beta_2 M_{it-1} \\
\mathrm{log}\sigma_{it} & = s_h(Y_{it-1})
\end{array} \right.
\end{align*}
\begin{align*}
\left\{ \begin{array}{ll}
\Delta Y_{it} & \sim N \left(\mu_{it}, \sigma_{it} \right) \\
\mu_{it} & = \alpha + \beta_1 Y_{it-1} + s_1(A_{it},t) + \beta_2 M_{it-1} s_2(A_{it},t) \\
 & = \alpha + \beta_1 Y_{it-1} + (1 -M_{it-1}) s_1(A_{it},t) + \beta_2 M_{it-1} + M_{it-1} s_2^{\star}(A_{it},t) \\
\mathrm{log}\sigma_{it} & = s_h(Y_{it-1})
\end{array} \right. 
\end{align*}
These models correspond to the Directed Acyclic Graph shown in Figure \ref{fig.DAG.simple}. By including both a main effect for HAS and a smooth interaction between HAS, Age and Period, these models are not identifiable unless additional constraints are imposed. For $s_1$ and $s_2^{\star}$ the smooth functions are constrained such that their mean is zero. $s_2$ must be constrained to have mean equal to one as is typical for a Varying Coefficient Model. Albeit both formula containing two two-dimensional tensor splines are mathematically equivalent, they may result in different estimates depending on how the knots are chosen and how penalties are implemented. In this paper we used the second form only.

We expect that models including smooth terms will result in patterns that may be of interest. Unlike the entry and exit models, additional covariates are included in models. To calculate an average effect of exposure, we use the \emph{potential outcomes} framework \citep{rubin1974estimating,holland1986statistics,hernan2010causal} to estimate the expected change in MH for each individual using their observed covariates and comparing these with the predicted outcomes for the counterfactual state wherein HAS is altered but all other covariates are retained. The average effect of exposure can be obtained by averaging the difference between predictions for HAS=1 minus the predictions for HAS=0. Performing this calculation, we must make the strong causal assumptions (1) that there are no (un)measured variables that affect the outcome other than the variables included in the model: age, calendar-year, previous Mental Health and HAS, (2) that exposure could have been different, and (3) that changing the exposure for one person does not affect exposure for other people. These assumptions are also referred to as exchangeability, manipulability and the stable unit treatment value assumption. If any of these assumptions is violated, the calculations do not provide a meaningful answer to a causal question.

Averaging the differences in counterfactual outcomes over all observed covariates provides an estimate of the average effect of HAS. We define the domain-specific average effect of HAS by as the domain-specific mean over all remaining variables, in essence over the observed values of previous MH. As all the models we compared are additive in previous MH and no transformation of the outcome was performed, the effect of HAS does not depend on previous MH and we do not need to average over all values of previous MH. Instead, we counterfactually set previous MH equal to any arbitrary value, e.g. its mean value, 75. Given a model that was fit using Markov Chain Monte Carlo, it is straightforward to obtain a sample of the posterior  for the potential outcomes and for the difference in potential outcomes between the counterfactual states of exposed and unexposed.\footnote{Using built-in functions from the package {\tt brms} to obtain a posterior sample of the linear predictor for each potential outcome, this amounts to the elementary operations subtraction and averaging.} As previous MH drops out of the equation for the differences, this allows graphing the expected domain-specific effect of HAS by age and calendar-year. To visualise the expected change in MH by age and calendar-year within each HAS group, we need to explicitly condition upon previous MH or average over it. In this case study the difference between the two options is an additive constant, however, this does not hold for non-linear models.

\subsection{Results}

We compare the performance of the models using 5-fold CV as we did for the entry and exit models: we calculate the expected log posterior predictive density out of sample and perform the cross-validation such that approximately twenty percent of each age-calendar-year and HAS stratum is included in each fold. We did not stratify on previous MH when allocating the folds.

Table \ref{tab.elpd.mh} summarises the predictive performance. The models can be grouped two ways. First, we compare the results between complete pooling, partial pooling and tensor splines. As expected, partial pooling and tensor splines perform better than complete pooling, with tensor splines providing the best results. This observation is true within the set of models without HAS, for the models including a main effect for HAS and also for the models allowing an interaction between age-period and HAS. 

Second, comparing the three models with partial pooling, allowing variation in the effect of HAS by age and calendar-year performed better than the model assuming no modification and the model assuming no effect of HAS. The same is true for tensor splines and the completely pooled estimate also performs worse than the model that includes a main term for HAS but no effects for age or calendar-year. For the latter model, domain-specific estimates of expected change in Mental Health, marginalised over HAS, do vary across the domains as HAS prevalence varies across the domains. Such marginal estimates can be considered a \emph{synthetic} estimator for the effect of age and time on Mental Health with HAS as an auxiliary variable if the effect of HAS is not of interest in itself.

Direct comparisons between models that differ both in modelling effects of HAS and in pooling/smoothing is non-trivial. For instance, complete pooling with a main effect of HAS does not perform better than tensor splines without a main effect of HAS. This does not imply that heterogeneity over the domains is more important than the effect of HAS. Let us presume that the prevalence of HAS varies across the domains and that HAS has an effect on Mental Health but there is no direct effect age or calendar-year on Mental Health. Marginalised over HAS, there will be heterogeneity in the domain-specific expected changes in Mental Health and this will be `detected' by a tensor spline model that does not include a term for HAS. The same is true if we assume both an effect of HAS and a direct effect of the domains. In essence, variation in the prevalence of HAS over the domains can be absorbed into the heterogeneity of the Mental Health changes over these domains. Conversely, comparison between the tensor splines only-model and a model including only HAS, does not provide a clear answer whether the observed heterogeneity was purely due to heterogeneity in HAS prevalence.

To directly assess whether HAS has an effect, the posterior estimates for its main term must be considered. Table \ref{tab.estimates} shows that results for the main terms are similar between the completely pooled and the tensor spline model with effect modification. Both models indicate an average loss of two to three points in Mental Health each year a person suffers HAS, however, the negative coefficient for previous Mental Health implies a regression to the mean: people with lower Mental Health scores improve over time, whereas people with near-perfect scores tend to score worse the next year. As such, continuous exposure to HAS results in a shift in equilibrium by -7.24 points: expected Mental Health is changed by -2.39 in the first year, an extra -2.39 -2.39*-0.33 in the second year \ldots{} until the negative pressure from HAS equals the pull towards the mean. This occurs when $\beta_2 + \beta_1 \Delta Y_{it-1} = 0$ or $\Delta Y_{it-1} = - \beta_2 / \beta_1 $. Using the estimates from the completely pooled model, this results in a shift of $-7.24 =-(-2.39)/(-0.33)$. Compared to the average and maximum Mental Health score, 75 respectively 100, this is a clinically important effect. 

\begin{table}[bt]
\caption{\label{tab.elpd.mh} Comparison of out of sample predictive performance of estimates of age-calendar-year-specific changes in Mental Health, using 5-fold Cross-validation stratified by age-calendar-year and HAS.}
\centering
\begin{tabular}{lrlrl}
\hline
Estimator & elpd & sd & $\Delta$ elpd & sd \\
\hline
 Complete pooling &  &  &  &  \\ 
   $\quad \mu_i = \mu + \beta_1 Y_{it-1}$ & -441283.6 & 388.9 & -240.1 & 23.3 \\ 
  Partial pooling &  &  &  &  \\ 
   $\quad \mu_i = \mu + \beta_1 Y_{it-1} + u_{at}$ & -441230.9 & 389.0 & -187.4 & 22.1 \\ 
  Tensor spline &  &  &  &  \\ 
   $\quad \mu_i = \mu + \beta_1 Y_{it-1} + s_1(at)$ & -441157.1 & 389.2 & -113.6 & 17.1 \\ 
  Complete pooling with HAS &  &  &  &  \\ 
   $\quad \mu_i = \mu + \beta_1 Y_{it-1} + \beta_2 X$ & -441171.0 & 388.1 & -127.5 & 15.9 \\ 
  Partial pooling with HAS &  &  &  &  \\ 
   $\quad \mu_i = \mu + \beta_1 Y_{it-1} + \beta_2 X + u_{at}$ & -441122.8 & 388.2 & -79.3 & 14.4 \\ 
  Tensor spline with HAS &  &  &  &  \\ 
   $\quad \mu_i = \mu + \beta_1 Y_{it-1} + \beta_2 X + s_1(at)$ & -441052.1 & 388.5 & -8.6 & 4.4 \\ 
  Partial pooling with HAS and modification &  &  &  &  \\ 
   $\quad \mu_i = \mu + \beta_1 Y_{it-1} + \beta_2 X + u_{atX}$ & -441116.9 & 388.3 & -73.4 & 14.2 \\ 
  Tensor spline with HAS and modification &  &  &  &  \\ 
   $\quad \mu_i = \mu + \beta_1 Y_{it-1} + \beta_2 X + (1-X) s_1(at) + X s_2(at)$ & -441043.5 & 388.5 &  &  \\ 
  
\hline
\end{tabular}
\end{table}

\begin{table}[bt]
\caption{Estimated effect of HAS and regression to the mean. \label{tab.estimates}}
\centering
\begin{tabular}{lcc}
\hline
& \multicolumn{2}{c}{posterior mean [95\% CI]} \\
Variable & Complete pooling  & Tensor spline with modification \\
\hline
$\hat{\beta_1}$ ($Y_{it-1}$)  & -0.33 [-0.34; -0.33] & -0.33 [-0.34; -0.33] \\
$\hat{\beta_2}$ (HAS)  & -2.39 [-2.70; -2.08] & -2.85 [-3.73; -2.03] \\
\hline
\end{tabular}
\end{table}

\begin{figure}
\centering
\includegraphics[width=0.45\textwidth]{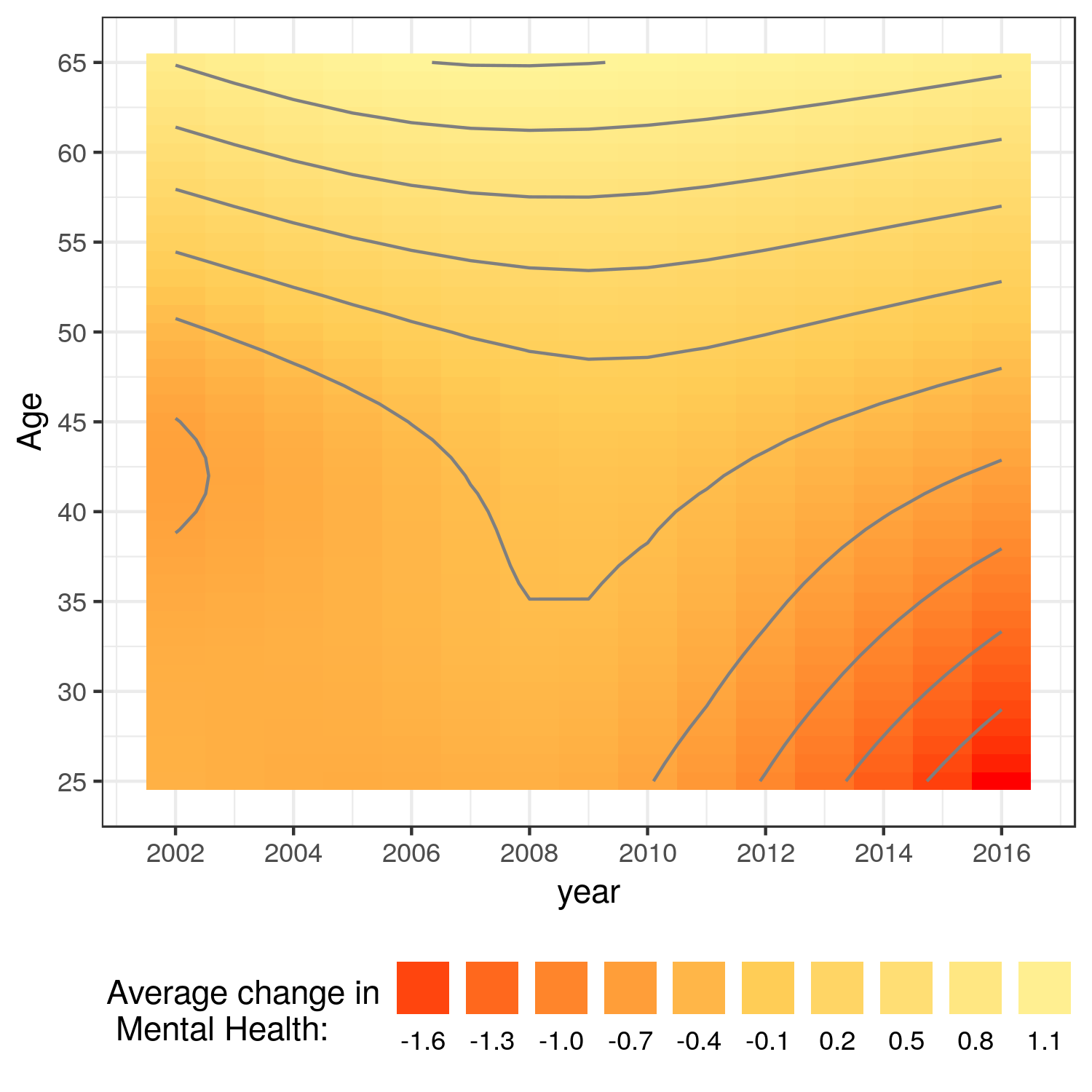} \quad 
\includegraphics[width=0.45\textwidth]{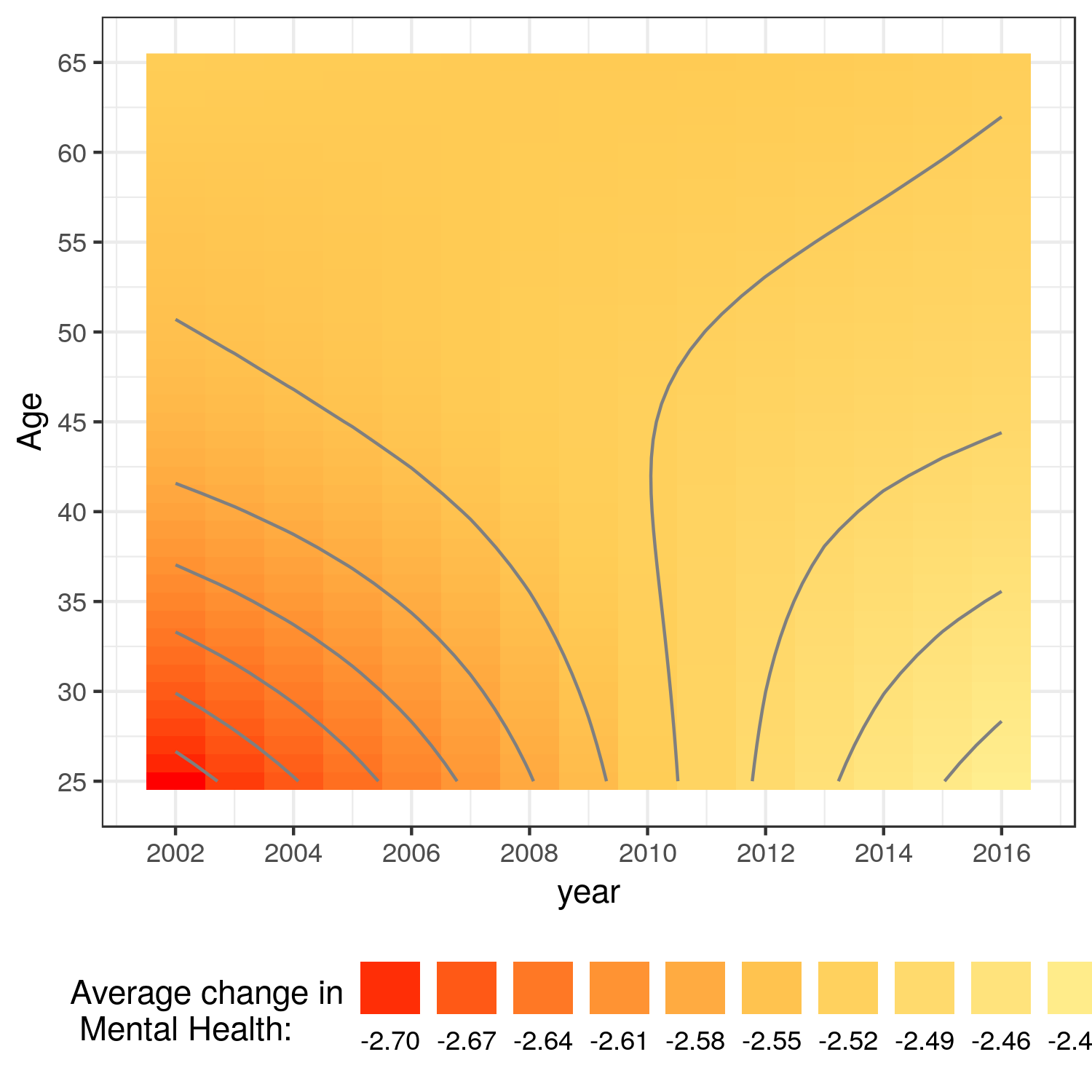} \\
\includegraphics[width=0.45\textwidth]{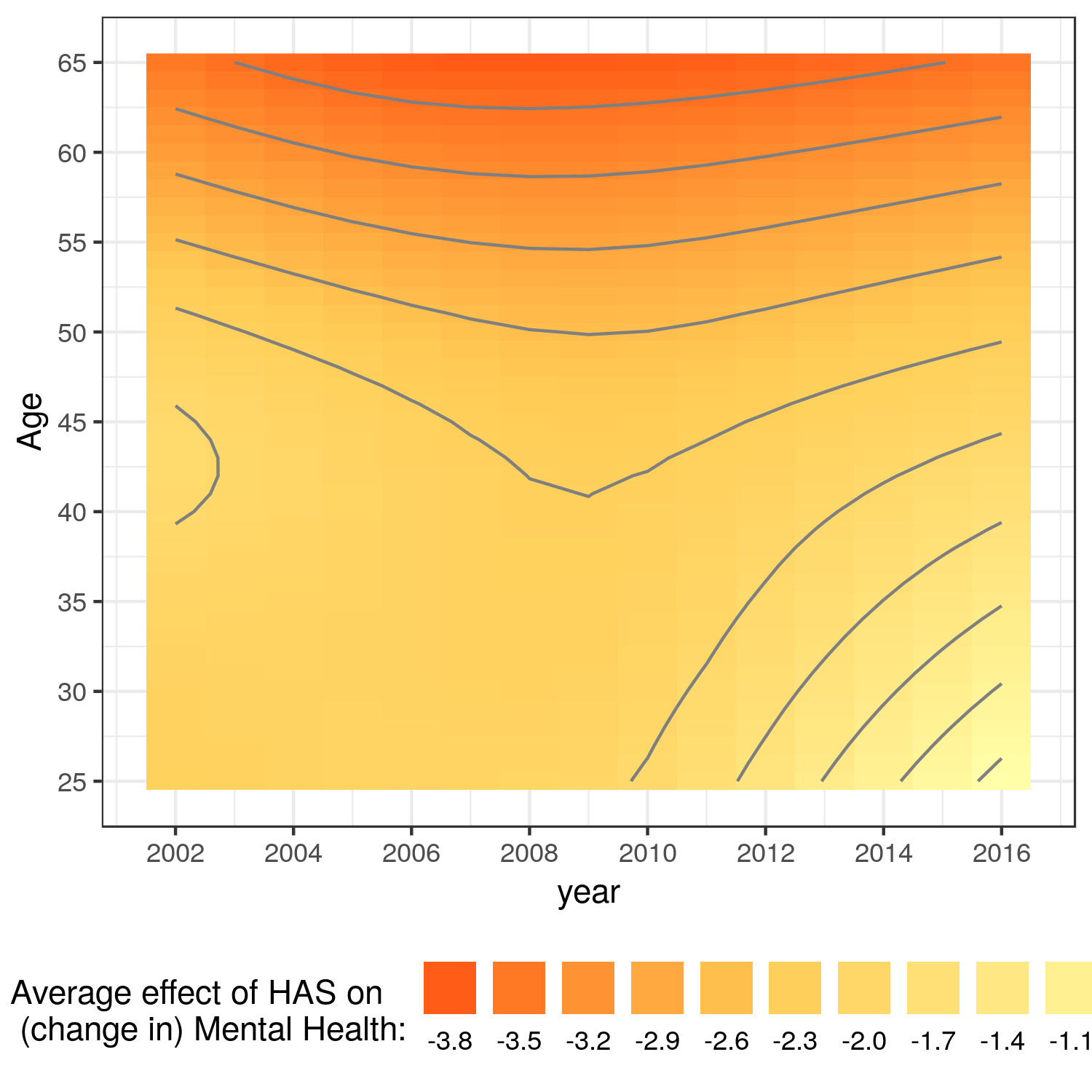} \quad
\includegraphics[width=0.45\textwidth]{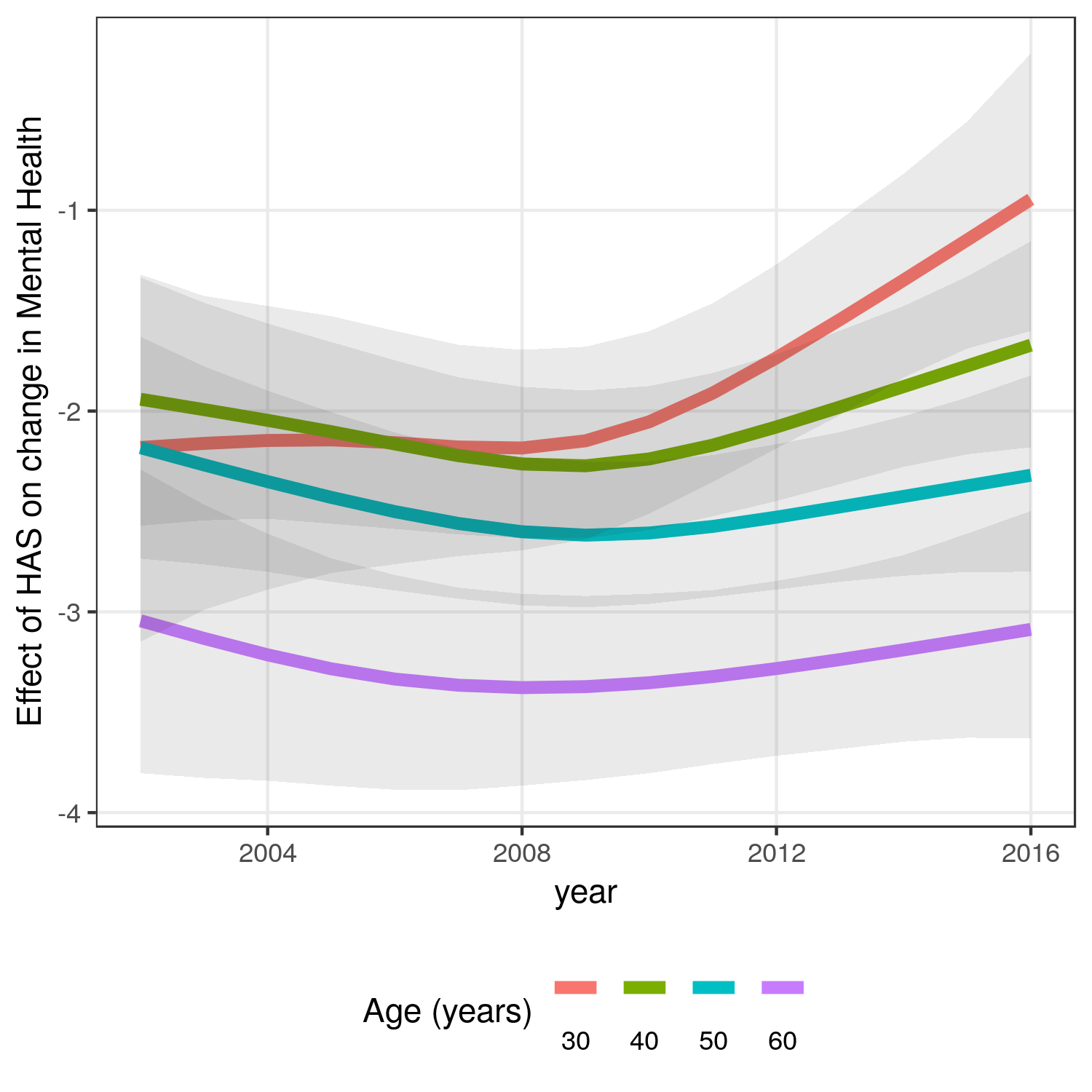}
\caption{Predicted change in mental health modelled using Tensor splines, allowing for a modification of the effect of HAS by age and time.\\
Top left: expected change in mental health for people the HAS-free group with a previous MH score of 75.\\
Top right: expected change in mental health for people the HAS group with a previous MH score of 75.\\
Bottom: average effect of HAS on change in mental health. \label{fig.mhtensorsmooth}}
\end{figure}

\begin{figure}
\centering
\includegraphics[width=0.45\textwidth]{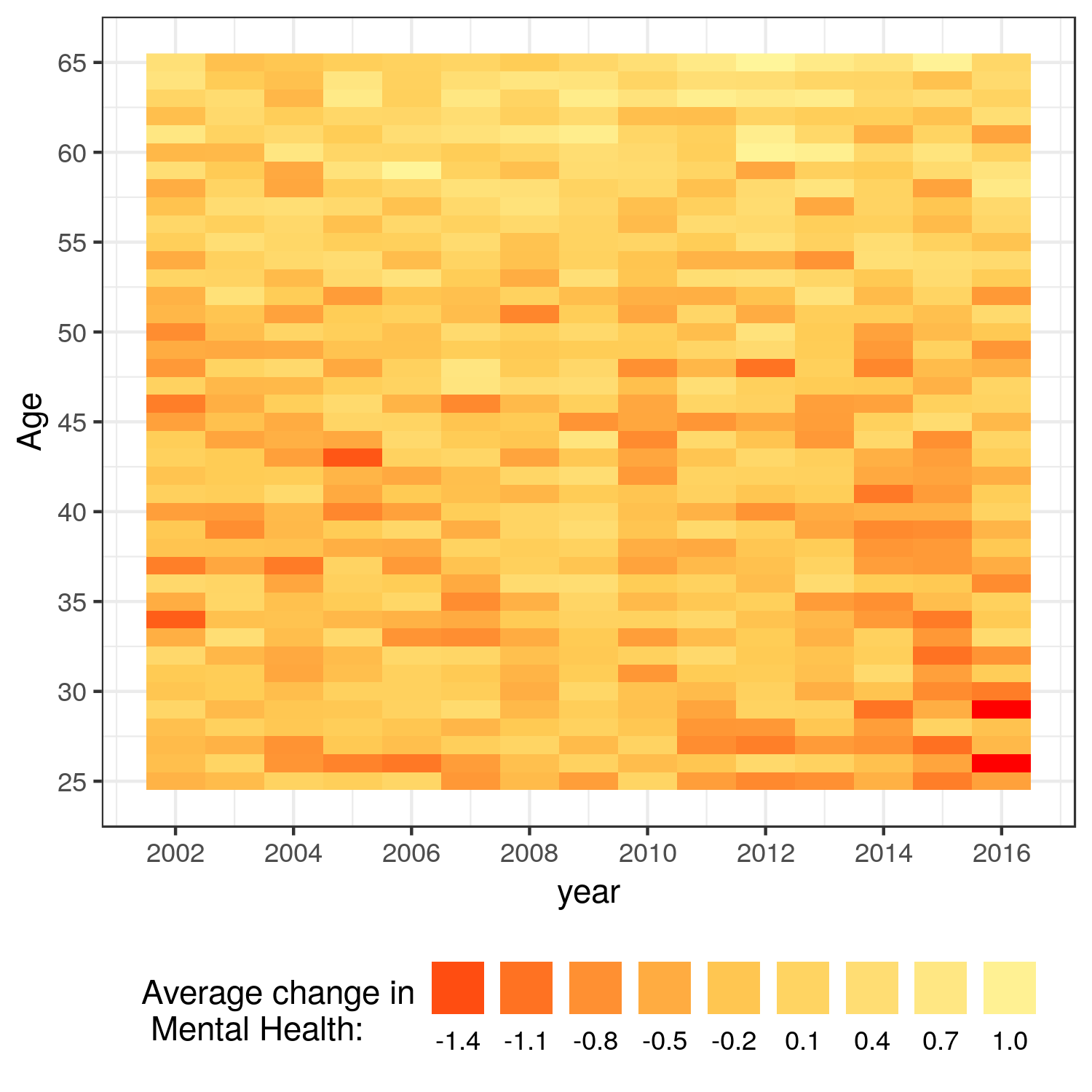} \quad 
\includegraphics[width=0.45\textwidth]{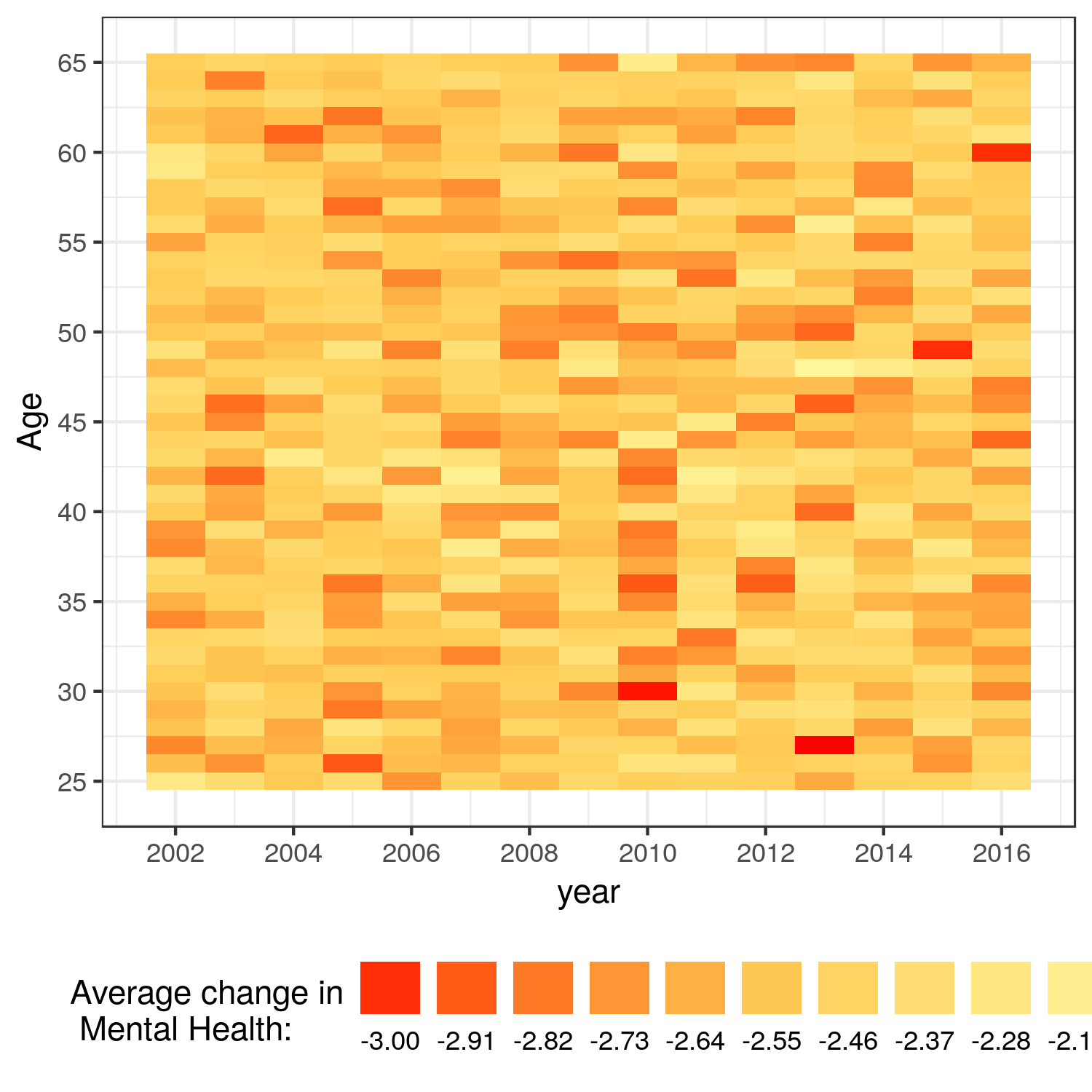} \\
\includegraphics[width=0.45\textwidth]{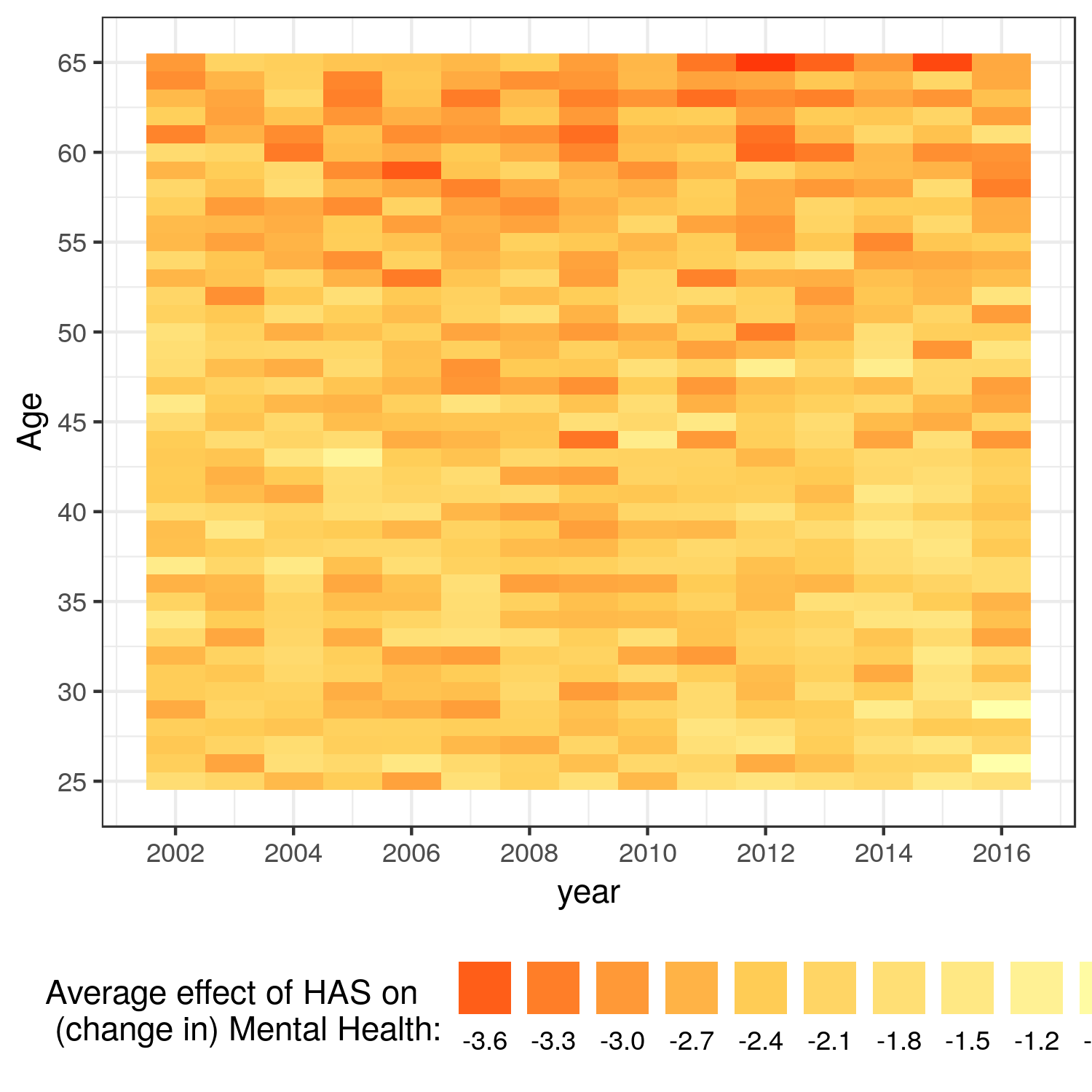} \quad
\includegraphics[width=0.45\textwidth]{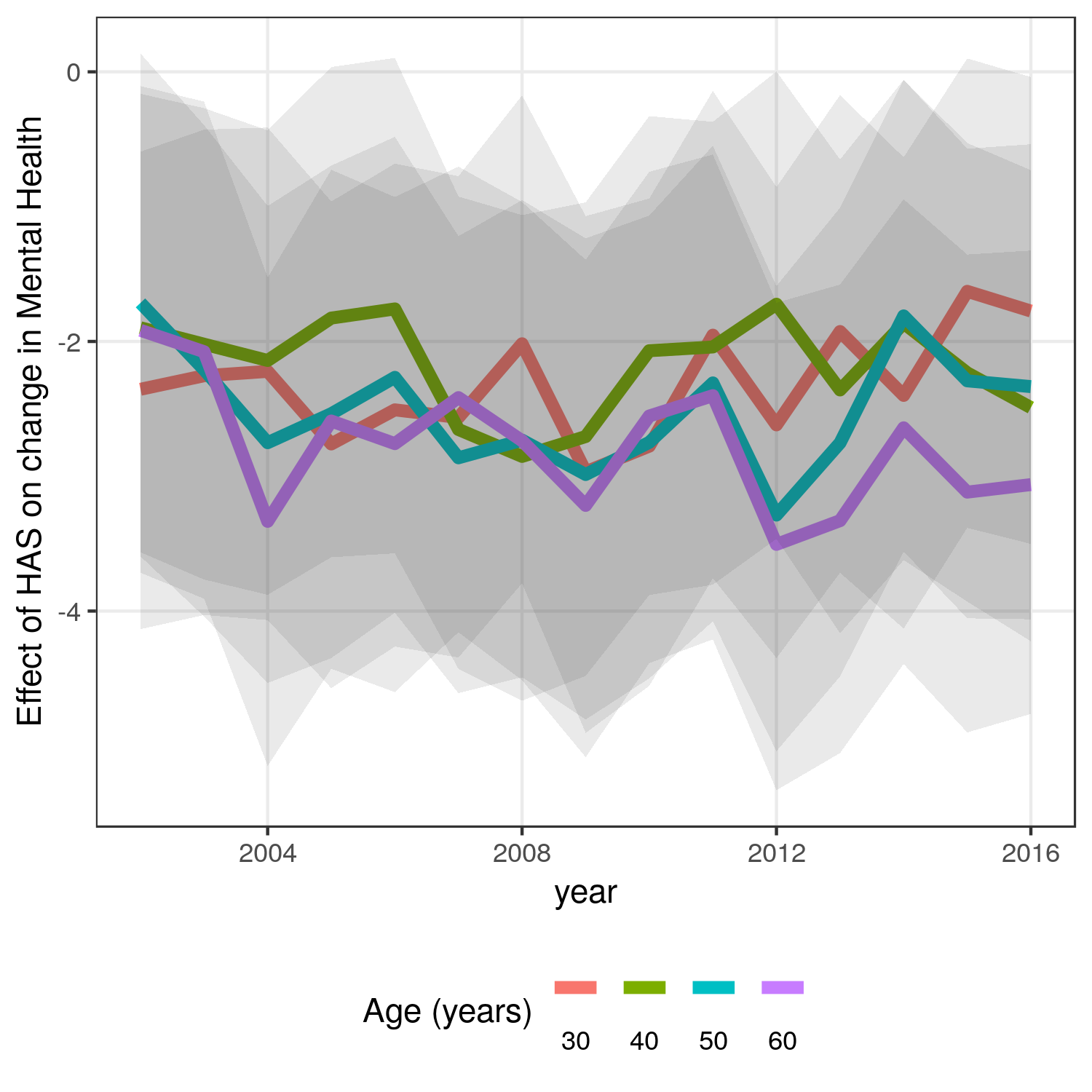}
\caption{Predicted change in mental health modelled using partial pooling, allowing for a modification of the effect of HAS by age and time.\\
Top left: expected change in mental health for people the HAS-free group with a previous MH score of 75.\\
Top right: expected change in mental health for people the HAS group with a previous MH score of 75.\\
Bottom: average effect of HAS on change in mental health. \label{fig.mhpartialpool}}
\end{figure}

These results indicate that there is an effect of HAS and that there is heterogeneity of the effect of HAS over age and time. The top panels in Figures \ref{fig.mhtensorsmooth} and \ref{fig.mhpartialpool} show the estimated expected changes in Mental Health for people with an initial score of 75 for people in HAS and out of HAS. When tensor splines are used to calculate these counterfactual effects, patterns are apparent in both counterfactual exposures. Similarly, patterns are revealed in the bottom panel for the tensor spline-based estimates but not for partial pooling. These panels describe the pattern in the average effect of HAS as modified by age and calendar-year. The range of expectation of the estimates is -3.8 to -1.1, providing a difference in effect size of 2.7 between the least and most affected. This estimated level of heterogeneity is on par with the size of the average effect. The strongest gradient is in age, with older people being more affected by HAS than younger people. In addition, for younger people the effect is decreasing in magnitude since 2009. However, looking at temporal patterns for younger HAS and younger HAS-free people, it appears the change in Mental Health is relatively stable in the HAS group but worsening since 2009 in the HAS-free group.

\section{Discussion}

We have illustrated that SDE methods are useful to visualise age- and time-specific data and that SDE estimates have better accuracy than direct estimation and complete pooling. In our examples, tensor splines provided the most accurate estimates. \emph{By design} these estimates vary smoothly over age and calendar time; it should not be a surprise that assuming smoothness results in smooth estimates. Our preference for these estimates is based both on a prior belief that such patterns are plausible, the improved out of sample predictive performance and the way they facilitate hypothesis generation. Partially pooled estimates also outperformed complete pooling in 5-fold cross-validation and we prefer them over completely pooled estimates, however, partial pooling did not reveal patterns and the improvement was substantially lower than for tensor splines.

We offer an intuitive explanation why smoothing performs well when patterns exist: a smooth estimate for domain $(a,t)$ `borrows strength' only from those domains that are quite similar, e.g. $(a,t-1)$, and incur negligible bias if indeed $\theta_{at} \approx \theta_{at-1}$ and $\theta_{at} \approx \theta_{a-1t}$. In contrast, partial pooling `borrows strength' from all other domains. If the heterogeneity between the domains  contains patterns (gradients, clusters, autocorrelations) instead of white noise, then partially pooled estimates (and completely pooled estimates) are biased: $E[\hat{\theta}_{at}] \neq \theta_{at}$ but it is simultaneously possible for the average over the domain-specific estimates to be unbiased: $E[\sum_{at}\hat{\theta}_{at}] = \sum_{at}\theta_{at}$ as $\vert{} E [\sum_{at} \hat{\theta}_{at} - \theta_{at}] \vert{} = \vert{} \sum_{at} E [\hat{\theta}_{at} - \theta_{at}] \vert{} \leq  \sum_{at} \vert{} E[  \hat{\theta}_{at} - \theta_{at} ] \vert{}$. 

Clearly, the total absolute bias of partially or completely pooled estimates increases with the strength of the underlying patterns. Therefore, we expect similar empirical results whereever moderate or strong gradients exist. This can be examined post-hoc by looking at the amount of variation over the domain-specific estimates and by comparing models using cross-validation; comparing within-sample predictive performance is not recommended as this will favour the most flexible model. Cross-validation has previously been used to compare the predictive performance of direct estimators, partial pooling and complete pooling. \cite{wang2014difficulty} observed that partial pooling has lower prediction error than complete pooling regardless of sample size, is better than the direct estimator for small samples and performs similar for large sample sizes. Although they recommend cross-validation over in-sample training loss, they remark that `the improvement of the multilevel model as given by cross-validation is surprisingly tiny, almost negligible to unsuspecting eyes'. In our applications, the improvement was clear. Both the amount of heterogeneity and sample size played a role in this. Although we have not assessed this directly, it is intuitive that the difference between partial pooling and tensor splines depends on whether the heterogeneity across domains is dominated by white noise or by structure. A three-way comparison between complete pooling, partial pooling and tensor splines is recommended but further modification could be considered, e.g. if a discontinuity is suspected around the Global Financial Crisis in 2008, this directly opposes the assumption that 2008 is equally similar to 2007 as to 2009; the current approach could be extended by borrowing from interrupted time series approaches.

The analyses presented in this paper did not adjust for sex, educational status or region. Incorporating more information through auxiliary data into the model may improve the reliability of the estimates, and this is a standard procedure in SDE. For example, if we are not interested in the effect of sex {\em per se} but suspect that sex has a strong influence on the rate of exit, we could define a \emph{synthetic} estimator for the age-period domains that uses information on sex: $\hat{\theta}_{at}^{s} = w_{atM} \hat{\theta}_{atM} + w_{atF} \hat{\theta}_{atF}$. Although this estimator is based on smaller domains, it can be efficient, particularly if the effect of sex is relatively strong, additive on the linear predictor scale and if the ratio of men to women varied over age-calendar-year in the sample or in the target population. A similar model-based approach is \emph{Multilevel Regression and Poststratification} (MRP), which has recently become popular as a method to obtain estimates of population groups when the frequency of subgroups within the sample may differ from the target population(s). In this technique partial pooling is used to obtain small sub-domain estimates and subsequently these estimates are weighted to better represent their contribution to the target population in the domains of interest \citep{gelman1997poststratification,downes2019multilevel}. We did not use post-stratification in this manuscript but the same heuristic arguments apply to consider smoothing as an alternative to partial pooling for MRP.

{\bf APC models}

We proposed that the transition probabilities into and out of HAS are more similar between people of similar ages in the same or adjacent calendar-years, as opposed to people with many years of age or time between them. Equivalently, these probabilities are more similar between people of similar birth-years. Our choice to focus on Age and calendar-year (Period) was motivated by the structure of the data, which included people of working ages between 2001 and 2016. Although this covers multiple birth-years (Cohort), we did not observe these cohorts as part of the working population for equal periods of time. Nevertheless, {\bf{A}}ge, {\bf{P}}eriod and {\bf{C}}ohort are linked: $A = P - C$. For the purposes of forecasting and attribution of effects, many models have been proposed to disentangle APC-effects. Some of these models impose an additive structure or that only two out of the APC-trio have an effect. Tensor splines are not additive, however, so we decided to `ignore' the contribution of Cohort effects; we argue that the use of flexible smoothing splines foregoes the need to  include explicitly all three terms as long as additivity is not forced upon the estimates. Consider a model that is quadratic and additive in A, P and C. We can rewrite such a model as one that is quadratic in A and P:
\begin{align*}
y & = \alpha + \beta_1 A + \beta_2 A^2 + \gamma_1 P + \gamma_2 P^2 + \delta_1 C + \delta_2 C^2 + \epsilon \\
  & = \alpha + \beta_1 A + \beta_2 A^2 + \gamma_1 P + \gamma_2 P^2 + \delta_1 \left( P - A \right) + \delta_2 \left( P^2 - 2 A P + A^2 \right) + \epsilon \\
  & = \alpha + \left(\beta_1 - \delta_1 \right) A +  \left( \gamma_1 + \delta_1 \right) P + \left(\beta_2 + \delta_2 \right) A^2 + 
  \left( \gamma_2 + \delta_2 \right) P^2 - 2 \delta_2 AP + \epsilon
\end{align*}
Hence, removing C induces an `interaction' between A and P. This does not mean that an interaction between Age and Period proves that a Cohort effect must exist, unless such interaction can be excluded on other grounds. Similar results hold for cubic and higher order polynomials. Therefore, one of A, P and C can safely be ignored when using tensor splines to generate domain-specific estimates.\footnote{Attempts to fit additive models, $s_1(A) + s_2(P)$, were abandoned as the Markov Chain Monte Carlo procedure did not seem to converge.} All three need to be considered when interpreting the obtained patterns.

{\bf Forecasting and prediction}

Our main motivation for using SDE was to reveal patterns and aid in the generation of hypotheses. Technically, it is possible to obtain forecasts from the models using tensor splines. Cross-validating the predictive performance by leaving out entire calendar-years, demonstrates that tensor splines perform better than partial pooling and complete pooling. This out of sample result may seem promising, however, beyond the one-year-ahead forecast the uncertainty increases rapidly as demonstrated in Figure \ref{fig.exit.forecast} in appendix. Effectively, the forecast is similar to one obtained by linearly extrapolating the temporal trends seen within each age group but placing most weights on the most recent 20\% of the data. Our smoothing (and partial pooling) approach was not designed for the purpose of forecasting and is unlikely to perform better than a method explicitly tailored to this task. This is consistent with \cite{booth2008mortality}: `The APC model has been usefully applied in describing the past, but has been considered less useful in forecasting.' Nevertheless, specialised models such as the Lee-Carter model \citep{lee1992modeling} can benefit from adopting smoothing \citep{currie2013smoothing,de2006extending}. 

Fortunately, SDE is suitable for prediction within-sample. While this is not relevant for the estimation of entry or exit rates, prediction is an essential step in the estimation of the expected change in Mental Health. Our estimation of the domain-specific effect of HAS on Mental Health is a form of semi-parametric g-computation \citep{robins1986new,snowden2011implementation}: we predicted the (counterfactual) outcomes for each person using their real age and period but setting previous Mental Health to 75 and setting HAS first to one and then to zero. Subtracting the predicted outcomes provides an estimate of the domain-specific effect of HAS. With this approach we can use SDE methods to provide accurate domain-specific estimates of \emph{associations} and to \emph{describe} the heterogeneity of these estimates. By themselves, such estimates are not causal estimates. A \emph{causal interpretation} depends on more assumptions relating to (unmeasured) confounding, manipulability, absence of spill-over effects etc.

{\bf Confounding, exchangeability and heterogeneity}
 

Since the incidence and escape from HAS depends on age and time, the prevalence of HAS depends on age and time but other factors could vary too. For instance, educational levels can vary over time; it is possible that people today have higher educational levels, on average, than people several decades ago. It is possible that this is true both within the HAS-free and the HAS subgroups, and, it is possible that the average educational level has increased more in the HAS-free group than in the HAS group. If we further assume that education has an effect on Mental Health, then this by itself is enough to provide heterogeneity in expected change in Mental Health both over age, calendar-year and HAS status. No direct effect of HAS is needed, and, the estimated effects and heterogeneity in estimated effects may be entirely due to a lack of exchangeability when they are caused by (unmeasured) changes in educational levels. 

Alternative mechanisms can, of course, result in similar patterns to those described above. For instance, rapid changes in housing costs may have altered strategies regarding home ownership. Postponing purchase and taking on a smaller mortgage could be sound economic advise, yet work contrary to prior expectations and lead to increased stress in those who strive to remain HAS-free. This mechanism leads to variability in changes in Mental Health but does not require demographic change or change in observed entry and exit rates but relies on changes in market prices and behaviour. Indeed, we can imagine the change in purchase strategy to perfectly counteract the change in pricing, so that the same people would remain HAS-free regardless of time. These people would possess different assets with, possibly, different consequences regarding their satisfaction and Mental Health. Whether we have observed the purchasing strategies or not, this example mechanism is different from the education example. It does not suppose change in the distribution of an individual level covariate but instead supposes a change in market conditions that affects everyone. Instead of a confounding variable that acts directly on Mental Health, the market influences Mental Health through the purchase strategy as a mediator.

For both mechanisms, heterogeneity in Mental Health over age and time is explained as heterogeneity in covariates (education) or context (house prices) over age and time. Whenever the ultimate goal of the research is to design interventions, attributing heterogeneity to observable - and hopefully manipulable - covariates is ideal. The residual heterogeneity over the domains is both a nuisance and an opportunity to further explain the observed patterns. Tables \ref{tab.elpd.mh} and \ref{tab.estimates} demonstrate that models ignoring the heterogeneity have lower performance but provide narrower credible intervals for the effect of HAS. This increased precision is likely spurious; it is well known that under-estimating the intraclass correlation in multi-level models leads to overestimates of precision and hence it has been advocated to report estimates for the effect of interest and estimates of heterogeneity side by side \citep{merlo2009individual,merlo2018general}. 

{\bf Semi-parametric versus non-parametric smoothing}

\begin{figure}
\centering
\begin{tikzpicture}[
	align=center,node distance=2cm,
	squarednode/.style={rectangle, draw=black, rounded corners, thick, minimum size=5mm},
	roundnode/.style={ellipse, draw=black, thick, minimum size=5mm},
	scale=0.45
	]
	\node (sfdiff) at (347.89bp,234.0bp) [draw,ellipse,roundnode] {$\mathrm{Mental Health}_{t} - \mathrm{Mental Health}_{t-1}$};
  \node (sfn) at (435.89bp,162.0bp) [draw,ellipse,roundnode] {$\mathrm{Mental Health}_{t}$};
  \node (sfome) at (640.89bp,90.0bp) [draw,ellipse,squarednode] {$\mathrm{SF36}_{t-1}$};
  \node (afford) at (368.89bp,306.0bp) [draw,ellipse,squarednode] {$\mathrm{HAS}_{t}$};
  \node (sfdiffme) at (543.89bp,18.0bp) [draw,ellipse,squarednode] {$\mathrm{SF36}_{t} - \mathrm{SF36}_{t-1}$};
  \node (sfo) at (614.89bp,306.0bp) [draw,ellipse,roundnode] {$\mathrm{Mental Health}_{t-1}$};
  \node (U) at (597.89bp,162.0bp) [draw,ellipse,roundnode] {$U$};
  \node (apc) at (133.89bp,306.0bp) [draw,ellipse,squarednode] {$\mathrm{Age-Period-Cohort}_{t}$};
  \node (sfnme) at (446.89bp,90.0bp) [draw,ellipse,squarednode] {$\mathrm{SF36}_{t}$};
  \draw [->] (sfo) ..controls (622.23bp,263.22bp) and (629.5bp,218.46bp)  .. (633.89bp,180.0bp) .. controls (636.24bp,159.4bp) and (638.03bp,136.06bp)  .. node {$$} (sfome);
  \draw [->] (sfnme) ..controls (482.72bp,63.403bp) and (498.05bp,52.023bp)  .. node {$$} (sfdiffme);
  \draw [->] (U) ..controls (571.51bp,148.71bp) and (566.54bp,146.26bp)  .. (561.89bp,144.0bp) .. controls (538.81bp,132.81bp) and (512.96bp,120.65bp)  .. node {$$} (sfnme);
  \draw [->] (sfo) ..controls (516.5bp,279.47bp) and (464.23bp,265.37bp)  .. node {$$} (sfdiff);
  \draw [->] (afford) ..controls (361.32bp,280.05bp) and (358.61bp,270.77bp)  .. node {$$} (sfdiff);
  \draw [->] (sfn) ..controls (439.84bp,136.13bp) and (441.24bp,126.97bp)  .. node {$$} (sfnme);
  \draw [->] (apc) ..controls (214.24bp,278.97bp) and (253.4bp,265.79bp)  .. node {$$} (sfdiff);
  \draw [->] (sfdiff) ..controls (381.02bp,206.9bp) and (394.33bp,196.0bp)  .. node {$$} (sfn);
  \draw [->] (sfo) ..controls (616.26bp,267.77bp) and (613.5bp,235.81bp)  .. (595.89bp,216.0bp) .. controls (583.91bp,202.53bp) and (546.54bp,189.54bp)  .. node {$$} (sfn);
  \draw [->] (U) ..controls (613.12bp,136.49bp) and (619.33bp,126.1bp)  .. node {$$} (sfome);
  \draw [->] (sfome) ..controls (605.06bp,63.403bp) and (589.72bp,52.023bp)  .. node {$$} (sfdiffme);
\end{tikzpicture}
\caption{Direct Acyclic Graph for the evolution of Mental Health influenced by Housing Affordability Stress and an Age-Period-Cohort effect wherein SF36 is an imperfect measurement of Mental Health subject to time-invariant unmeasured personal characteristics.\label{fig.DAG.ME}}
\end{figure}
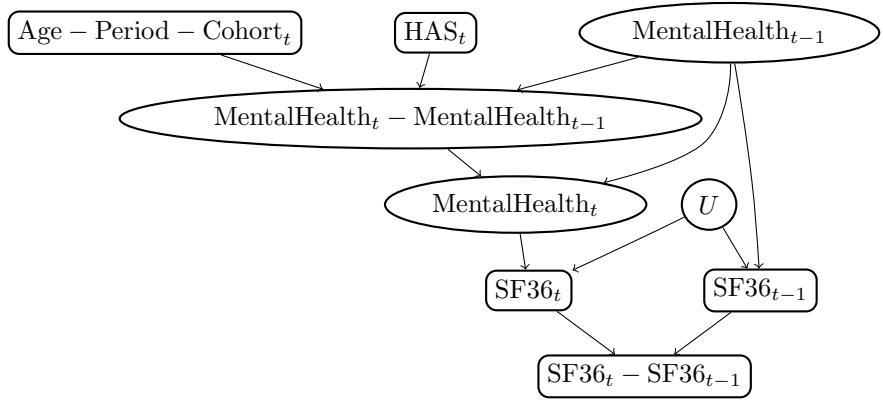

In this paper, we focussed on the technical aspects of obtaining precise estimates from a small number of observations. We modelled change in Mental Health as it is plausible that the change in Mental Health is independent of previous changes in Mental Health, whereas Mental Health itself is highly auto-correlated and likely to be a Markov Process. However, the Short-form 36 instrument has a limited precision and may be affected by time-invariant unmeasured characteristics as illustrated in Figure \ref{fig.DAG.ME}. While adjusting for previous Mental Health is conceptually sound, adjusting the previous SF36 score is not sufficient to remove potential unmeasured confounders $U$ \citep{glymour2005baseline}. It may seem that the `fixed effects model' is more appropriate as it explicitly aims to account for time-invariant confounders from the Mental Health measures, however, the standard fixed effects model assumes that conditional upon exposures, changes in outcomes are white noise; a `zero order' Markov process instead of a first order process. This encodes that losses in Mental Health are incurred immediately instead of accumulated over time, and, that they are reversed (forgotten) immediately when leaving HAS. The standard fixed effects model does not allow feedback loops. We refer to \cite{ding2019bracketing} and \cite{imai2019should} for further comparison between fixed effects (difference in difference) and lagged effects models.

We assumed a first order Markov Process and our model was additive in previous Mental Health. This allowed us to set previous Mental Health to 75 when predicting counterfactual outcomes as it factors out when subtracting the prediction for a person with HAS from the prediction for the same person without HAS. We could also have retained the observed previous Mental Health score and averaged the differences within each domain to provide an estimate of the marginal effect of HAS. As we've restricted the models to be additive in previous Mental Health, both estimates are identical but more complex scenarios can be imagined. For example, we could predict the outcomes if the exposure had been changed in 2001 and use the predicted outcomes as input to predict the subsequent exposure in 2002 etc. Indeed, the DAG in Figure \ref{fig.DAG.simple} could be extended to allow Mental Health as an input into the transition risks of subsequent HAS. This type of DAG was assumed by \cite{bentley2018impact}, who used marginal structural models to examine the effect of social housing versus private renting or ownership on Mental Health but did not target APC effects. Another line of research could be to investigate whether HAS is a mediator of the influence of `zeitgeist' on Mental Health and whether age, cohort and period operate mainly trough HAS or through other potential mediators.

In our application, smoothing provided a relatively small loss in stated precision compared to complete pooling and direct estimation, smaller than what could be expected from dividing the data into non-overlapping aggregates, e.g. 5-year age groups. As such, the smooth estimates allowed for a visualisation that revealed patterns and the existence of heterogeneity of effects. This approach is semi-parametric and can be compared with recent developments in non-parametric methods such as causal forests \citep{wager2018estimation}, which incorporate recursive partitioning into `honest' trees \citep{athey2016recursive}. Honest trees and causal forests are modifications of classification and regression trees, respectively random forests. These algorithms (repeatedly) partition the data into nodes that are homogeneous with respect to the treatment, the outcome or the treatment effect. By design each node is a convex region and can be seen as a small domain with the boundaries automatically chosen by the algorithm. In order to obtain unbiased (`honest') estimates of the effect, data must be split into a training set to grow the tree and a separate estimation dataset to estimate the treatment effect, which is different from using cross-validation to evaluate the accuracy of the procedure or to tune parameters to optimise the procedure. 

Causal forests have been used to estimate heterogeneity in treatment effects in the presence of a large number of covariates. In principle, the algorithm could also be applied to a two-dimensional problem and the resulting effect estimates could be visualised, potentially revealing trends or clusters similar to how tensor smoothing may show them. Causal forests do not use a penalty on wiggle but can be compared to weighted estimators. For each tree the domain $(a,t)$ must be included in the node that is used to estimate the treatment effect for this domain and other domains may contribute. The probability for other domains to be included depends on the structure of the data but neighbouring domains are more likely to contribute because the nodes must be convex regions. As a forest is an unweighted ensemble of trees, the final estimate for the effect in domain $(a,t)$ will be akin to a weighted estimate with most weight placed on $(a,t)$ and its neighbours; similar to a kernel that is adapted to both the data and the target domain. If the data generating process is a smooth function of $a$ and $t$, it is reasonable to expect that both our tensor splines and the causal forests will reveal patterns. However, causal forests are fully non-parametric, which foregoes the potential gains in efficiency by including knowledge or assumptions on structure such as the existence of conditional heterogeneity.

\section{Conclusion}

We have illustrated that tensor smoothing splines can be used to identify trends and hot spots in historical Age-Period-Cohort data. As a technique for small domain estimation it excels when such patterns exist, which is {\em a priori} not known. Cross-validation, stratified by domain, can be used to compare its performance with other estimators and provide evidence for the existence of this type of heterogeneity. 

Our second case study shows that effect modification by APC on a health effect of interest can be examined and that the magnitude of effect modification can be similar to the magnitude of the main effect. This information may help appreciate differences between studies: researchers often statistically control for age assuming that the effect of age is an additive confounder (nuisance) to remove from the estimate of interest. However, our results show that age may be a modifier of the effect and this may be important when comparing effect size estimates reported in different studies.

The combination of semi-parametric techniques suitable for small domain estimation with g-computation enables causal inference in the presence of strong effect modification and is of interest to develop targetted interventions. Such applications require support from a suitable Directed Acyclic Graph; by itself SDE remains useful as an aide to generate hypotheses.

\section*{Code}

The source code to generate tables and figures is available on \href{https://bitbucket.org/Koen_Simons/sde_for_apc/src/master/}{https://bitbucket.org/Koen\_Simons/sde\_for\_apc/src/master/}

\bibliographystyle{apalike}

\bibliography{smalldomain.bib}

\appendix

\section*{Additional Figures}

\begin{figure}
\centering
\includegraphics[width=0.45\textwidth]{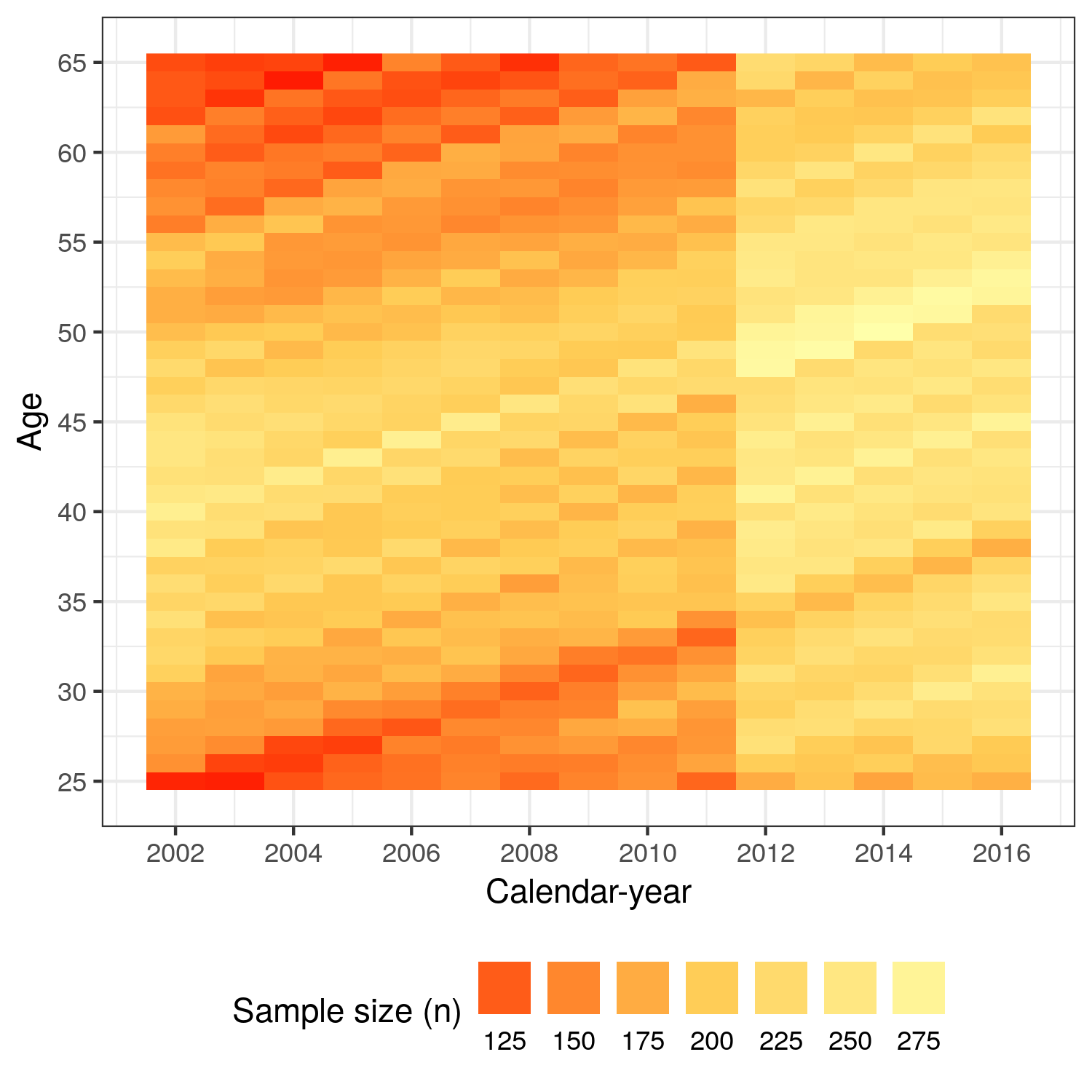} \quad \includegraphics[width=0.45\textwidth]{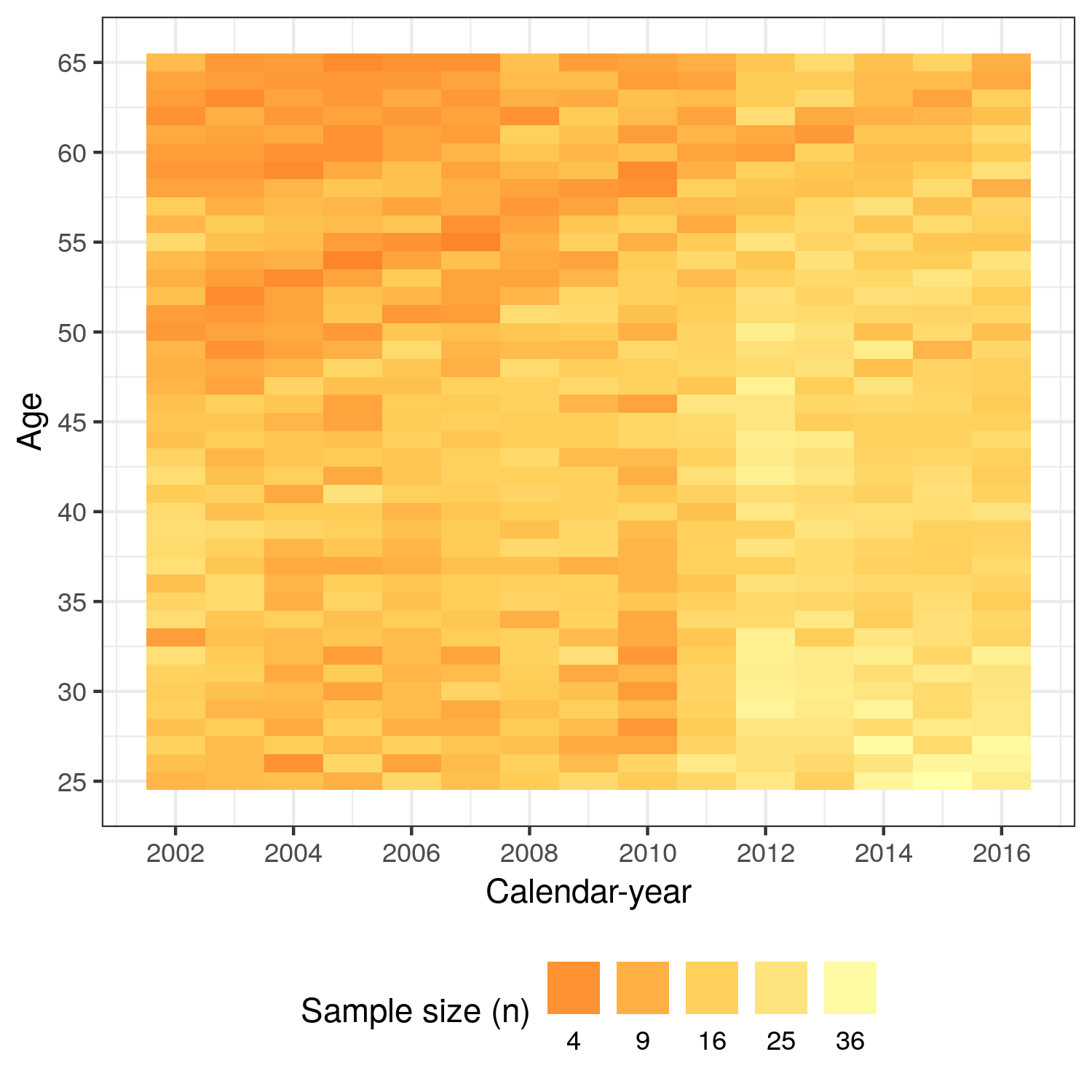}
\caption{Sample sizes. Domain-specific estimates using longitudinal data from the HILDA study must be obtained from relatively small sample sizes. Left: number of people without Housing Affordability Stress. Right: number of people in HAS. In both panels the sample size increases in 2012 due to a top-up sample. Diagonal lines appear due to people aging and remaining in the same group if they do not drop out of the study and do not enter into HAS or exit from HAS.\label{fig.atrisk}}
\end{figure}

\begin{figure}
\centering
\includegraphics[width=0.45\textwidth]{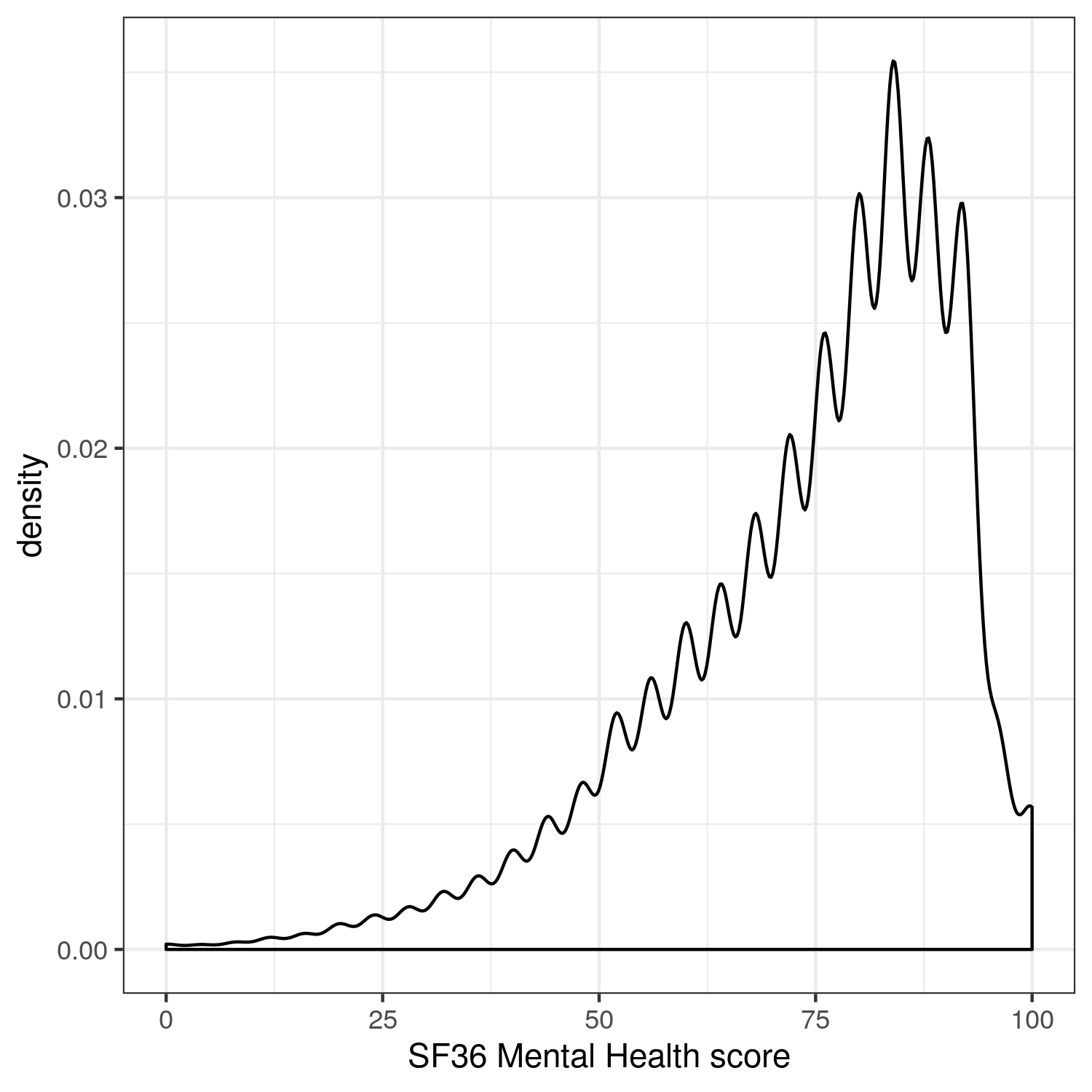} \quad \includegraphics[width=0.45\textwidth]{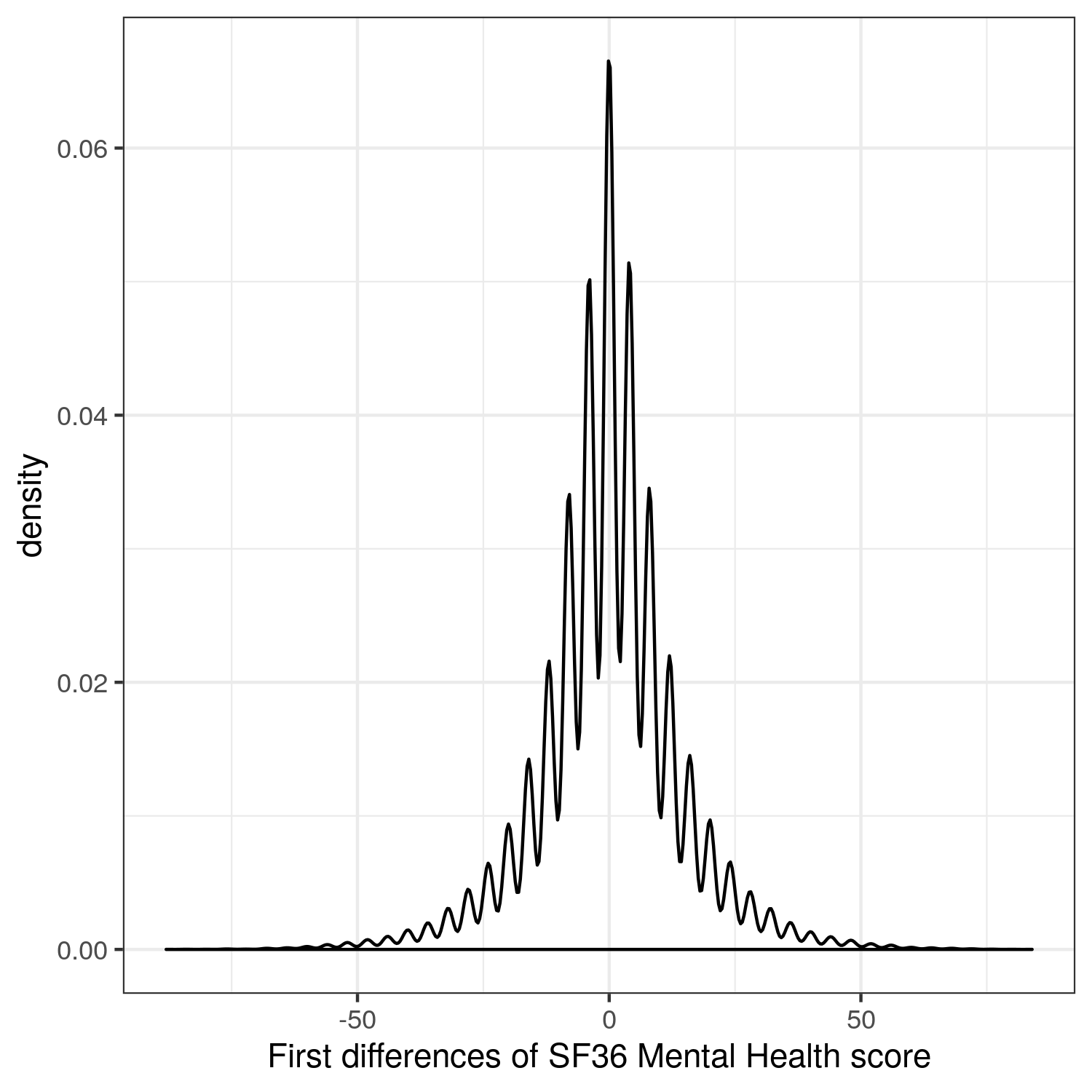} \\
\centering \includegraphics[width=0.45\textwidth]{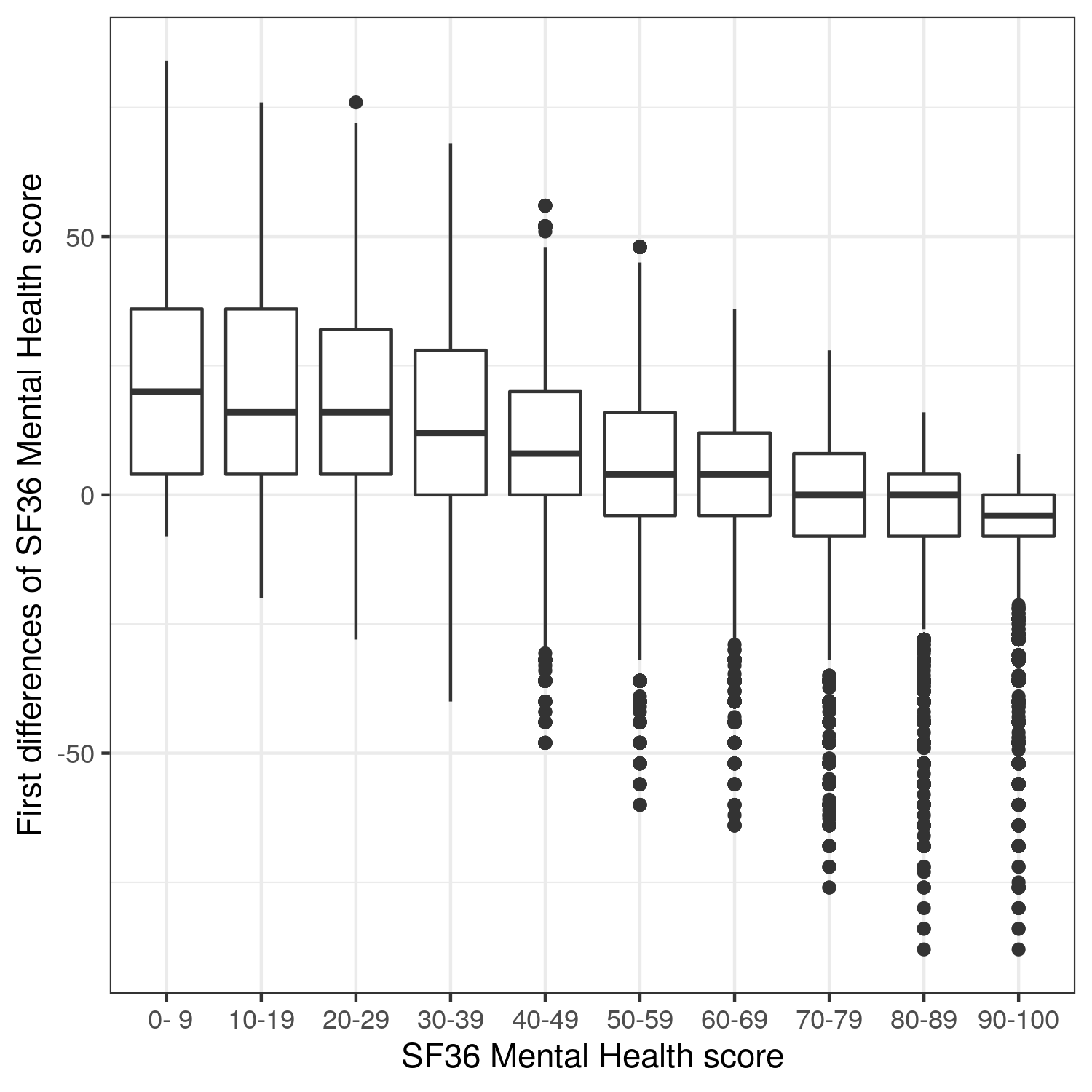} 
\caption{SF36 Mental Health scores. Top left: kernel density plot of the raw Mental Health scores, which have a skewed distribution. Top right: kernel density plot of the first differences of the Mental Health scores, which have a symmetric distribution. Bottom: First differences in Mental Health scores appear to be more variable for low previous Mental Health scores (conditional heterogeneity) and their means are negatively correlated with the previous score (regression to the mean).\label{fig.sf36}}
\end{figure}

\begin{figure}
\centering
\includegraphics[width=0.9\textwidth]{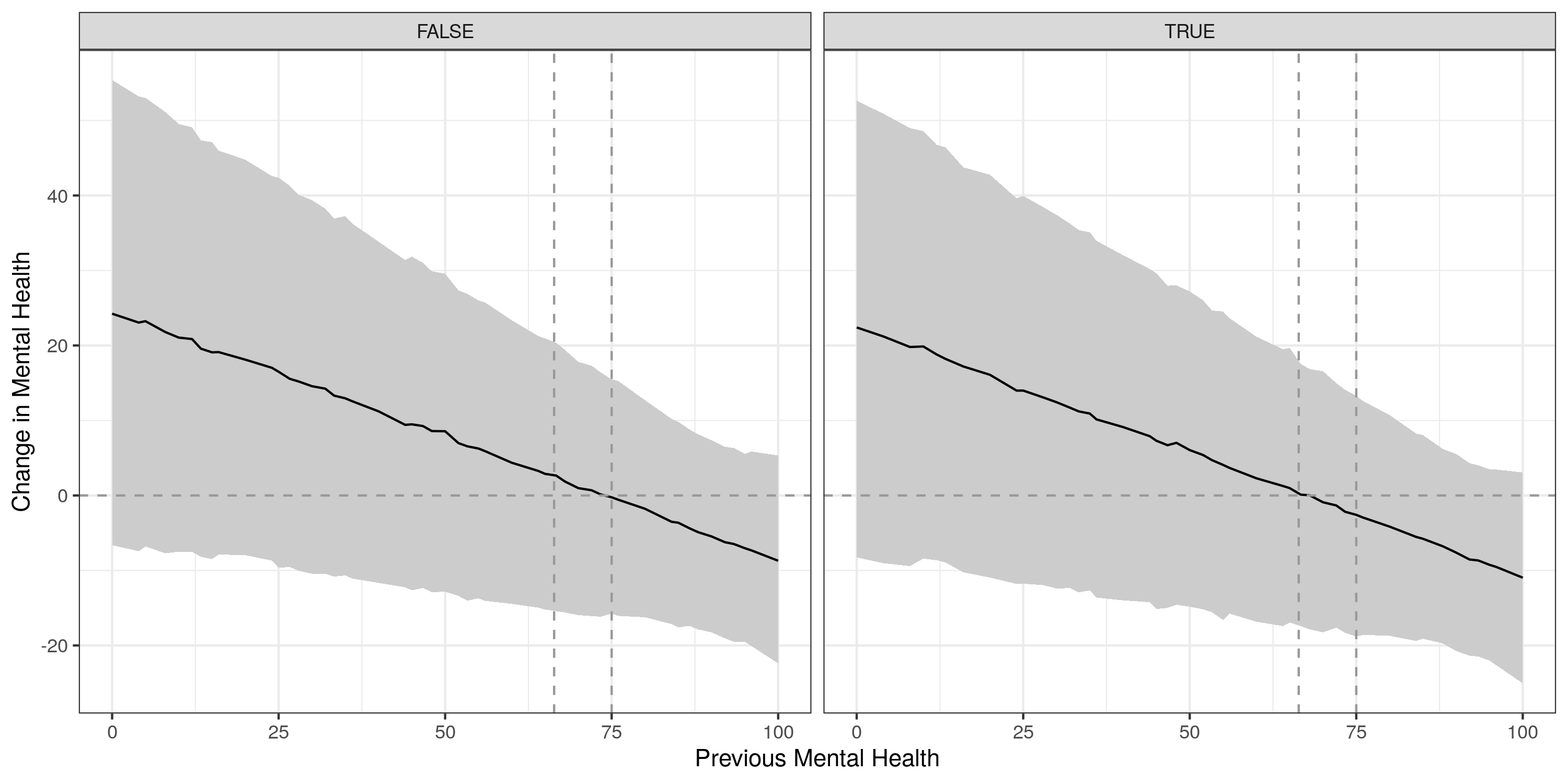}
\caption{Posterior estimates for the change in Mental Health as a function of previous Mental Health for age 50 in 2002, estimated using tensor spline smoothing and allowing for effect modification by age and period. Left: HAS-free people, Right: people in HAS. These graphs illustrate that the model allows both conditional heterogeneity (80\% pointwise CI is widest for lowest previous Mental Health scores), regression to the mean (negative correlation between change and previous score) and that the effect of HAS is a shift of the equilibrium equal to -7.24 points (=-2.85/0.33). \label{fig.conditionalheterogeneity}}
\end{figure}

\begin{figure}
\centering
\includegraphics[width=0.45\textwidth]{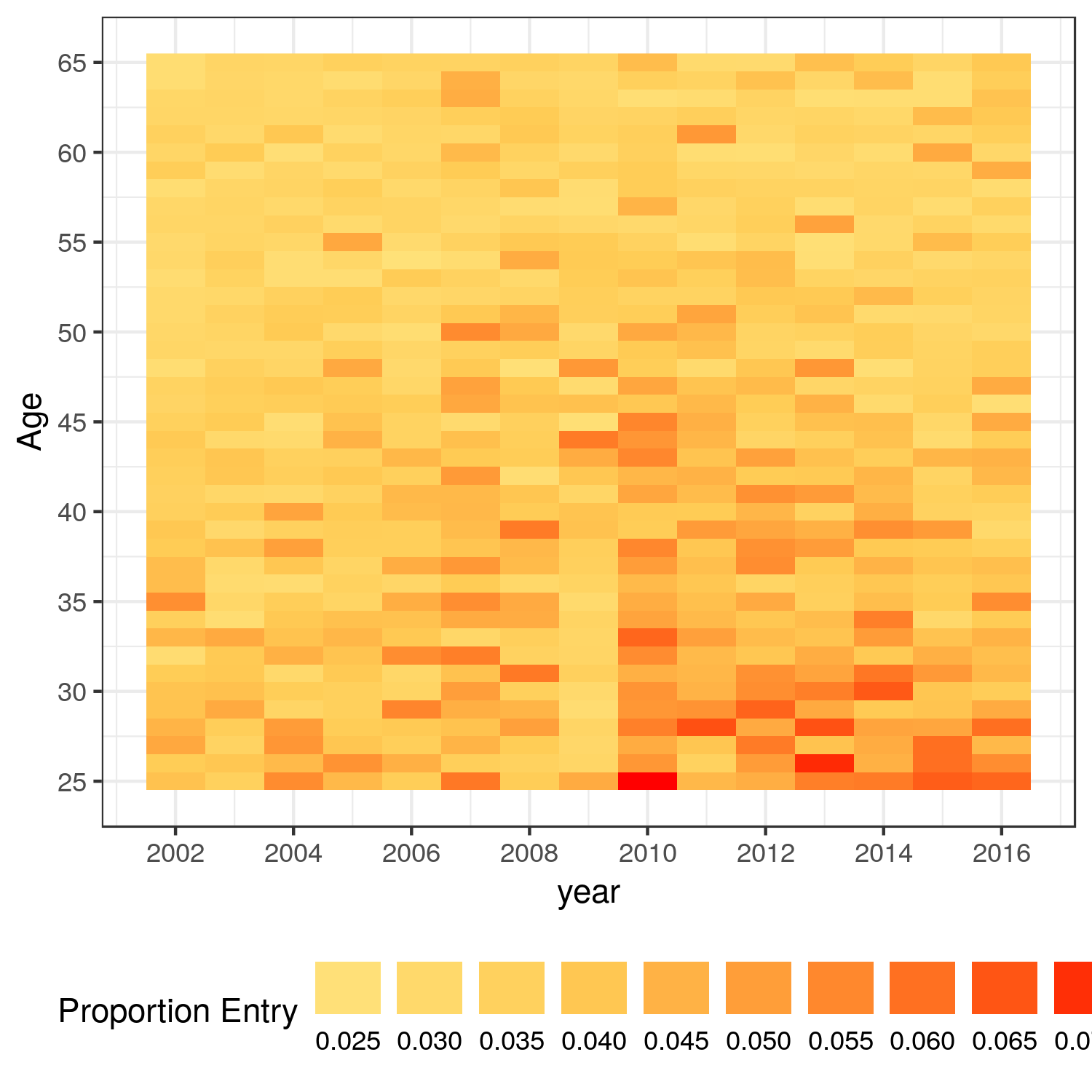} \quad \includegraphics[width=0.45\textwidth]{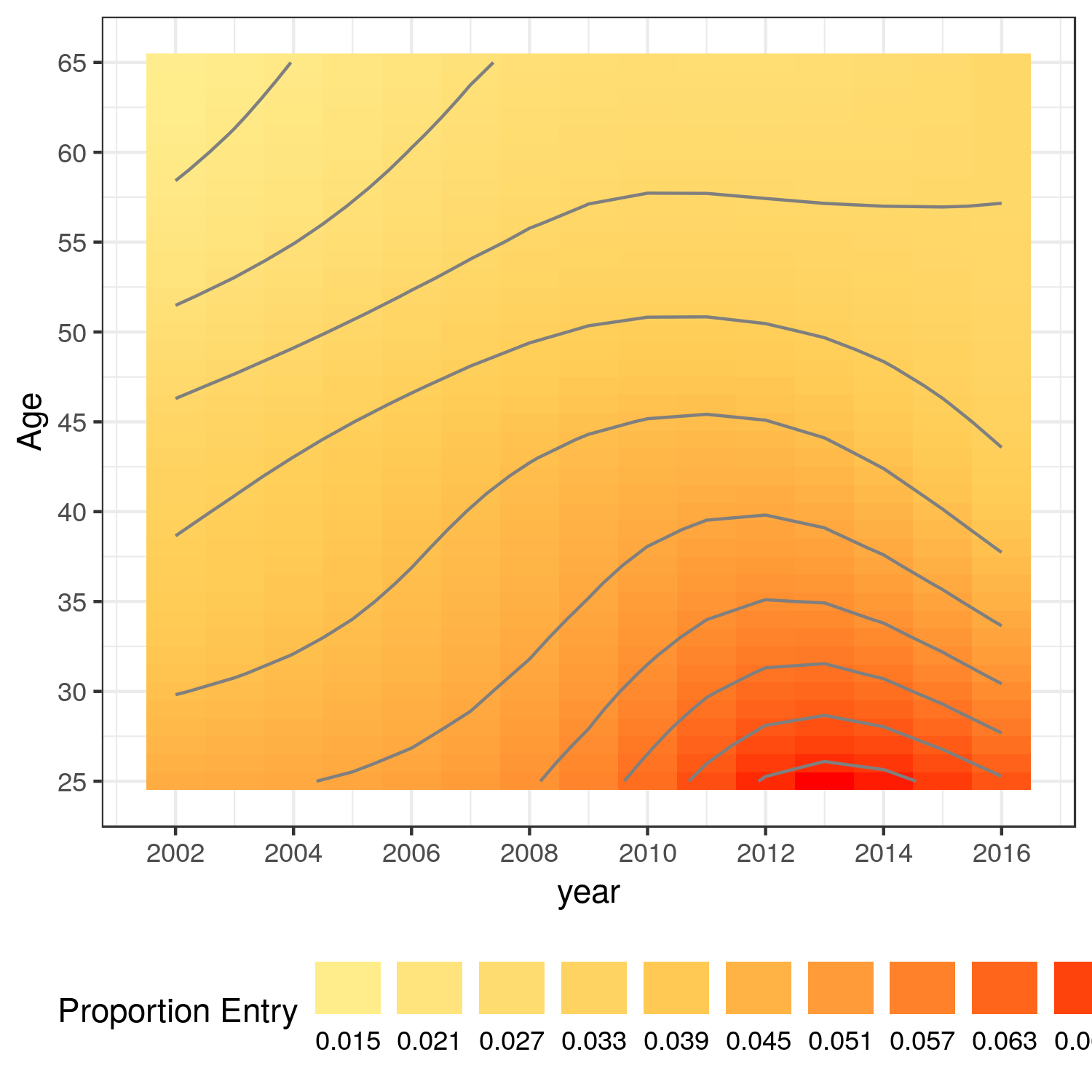} \\
\includegraphics[width=0.45\textwidth]{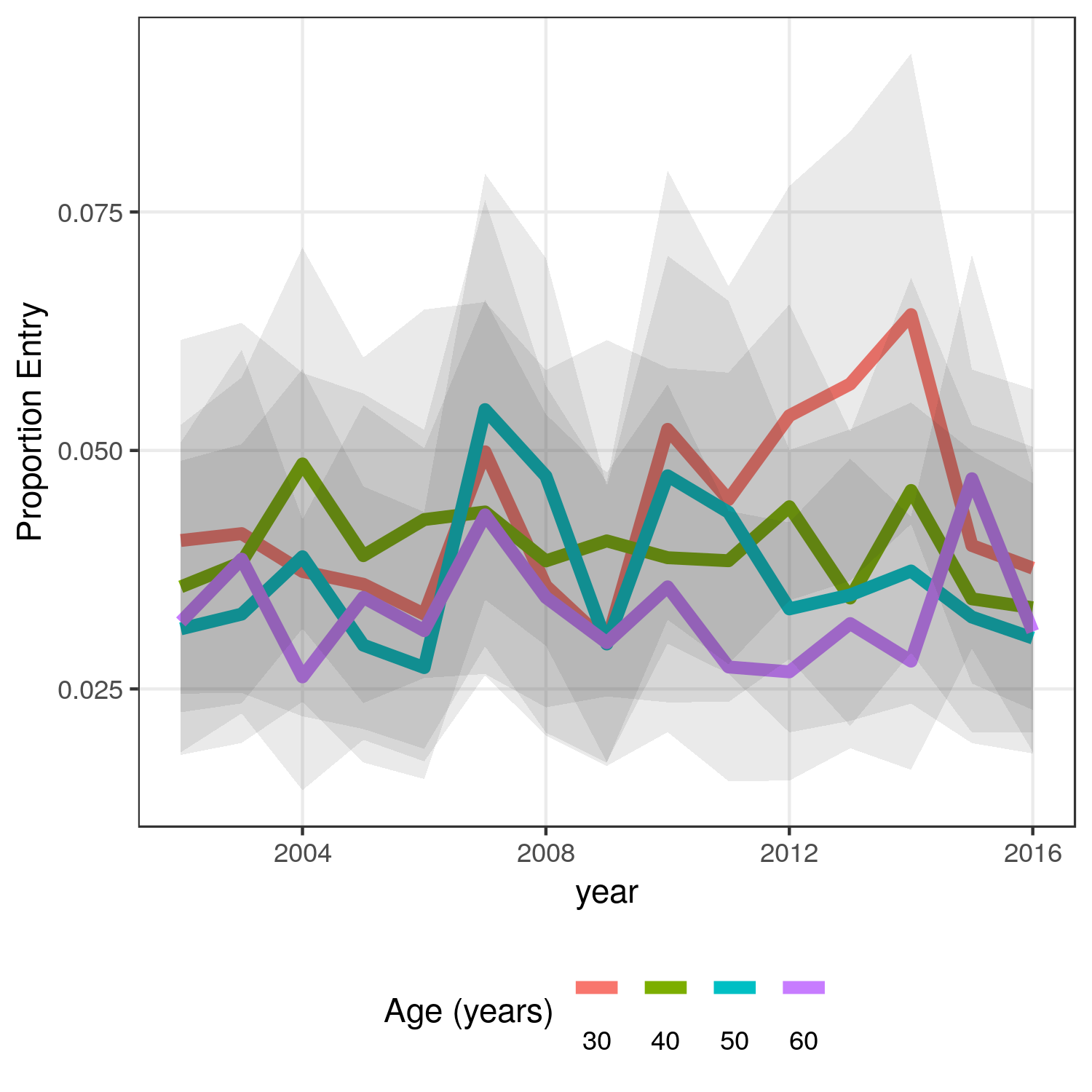}  \quad \includegraphics[width=0.45\textwidth]{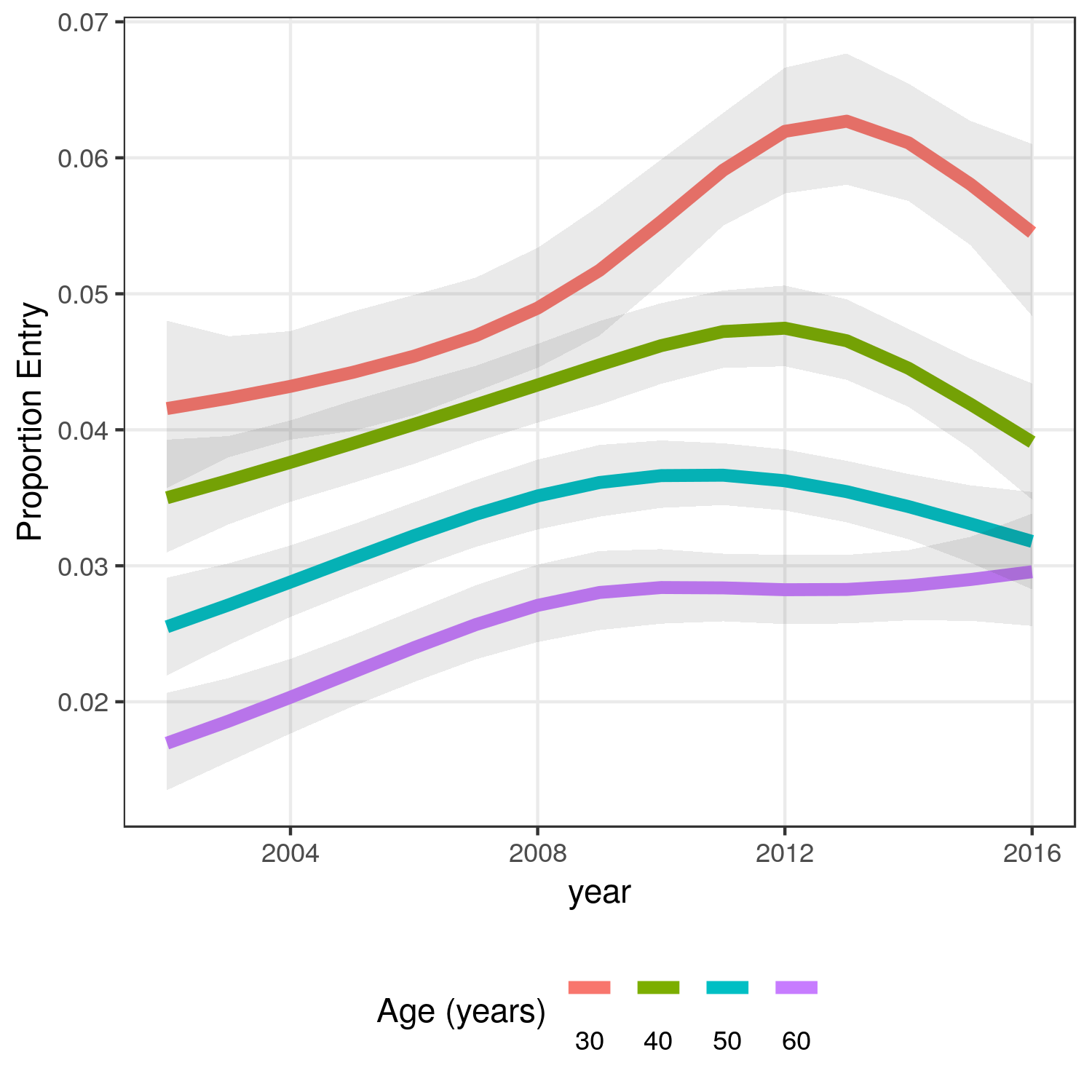}
\caption{Estimated probabilities of entering into Housing Affordability Stress by age and period. Top: Heatmaps of posterior means for the probability of exit by APC. Bottom: Distribution of the posterior for selected one-year age-groups. Left: map obtained from a partial pooling model with iid error. Right: map obtained using a tensor spline basis.\label{fig.entry}}
\end{figure}

\begin{figure}
\centering
\includegraphics[width=0.45\textwidth]{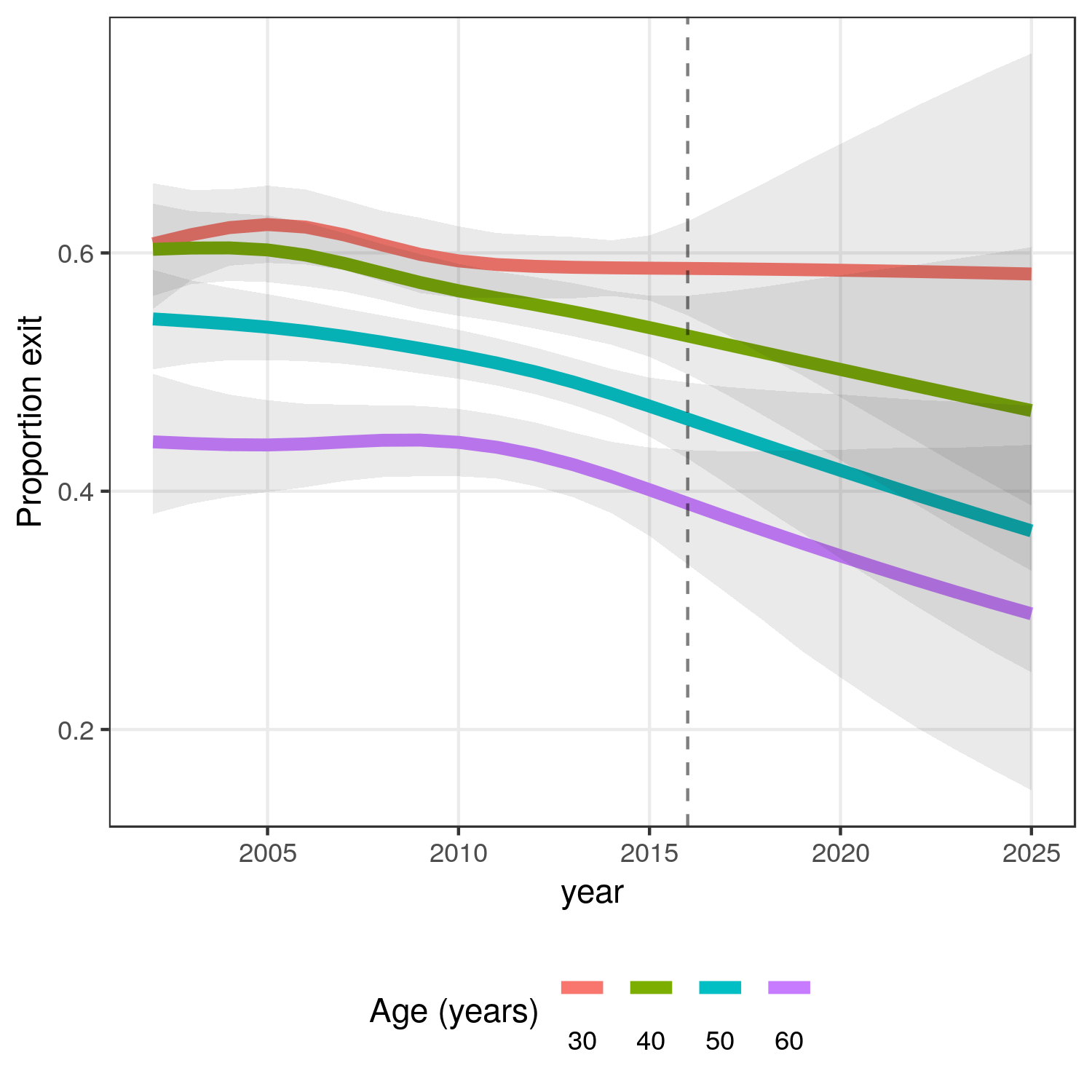} 
\caption{Forecasted probabilities of exit from Housing Affordability Stress by age. Forecasts obtained using tensor smoothing splines on age by period with five degrees of freedom on data spanning 2001 to 2016. Although a forecast can be obtained, it is similar to a linear extrapolation based on the most recent 20\% of the data and ignores potential cohort effects. Even with these limitations, the estimated precision decreases rapidly with the forecast horizon. \label{fig.exit.forecast}}
\end{figure}

\end{document}